\documentclass[aps,prb,superscriptaddress,twocolumn,floatfix]{revtex4}

\usepackage{graphicx}
\usepackage{dcolumn}
\usepackage{bm}
\usepackage{color}
\def\be{\begin{equation}}
\def\ee{\end{equation}}
\def\ba{\begin{array}{lll}}
\def\ea{\end{array}}
\def\ber{\begin{eqnarray}}
\def\eer{\end{eqnarray}}
\def\kv{{\bm k}}
\def\kvp{{\bm k}'}

\def\pv{{\bm p}}
\def\sigmav{{\bm \sigma}}
\begin{document}
\title{Theory of Coulomb drag for massless Dirac fermions}
\author{M. Carrega}
\affiliation{NEST, Istituto Nanoscienze-CNR and Scuola Normale Superiore, I-56126 Pisa, Italy}
\author{T. Tudorovskiy}
\affiliation{Radboud University Nijmegen, Institute for Molecules and Materials, NL-6525 AJ Nijmegen, The
Netherlands}
\author{A. Principi}
\affiliation{NEST, Istituto Nanoscienze-CNR and Scuola Normale Superiore, I-56126 Pisa, Italy}
\author{M.I. Katsnelson}
\affiliation{Radboud University Nijmegen, Institute for Molecules and Materials, NL-6525 AJ Nijmegen, The
Netherlands}
\author{Marco Polini}
\email{m.polini@sns.it}
\homepage{http://qti.sns.it}
\affiliation{NEST, Istituto Nanoscienze-CNR and Scuola Normale Superiore, I-56126 Pisa, Italy}
\begin{abstract}
Coulomb drag between two unhybridized graphene sheets separated by a dielectric spacer has recently attracted considerable theoretical interest. We first review, for the sake of completeness, the main analytical results which have been obtained by other authors. We then illustrate pedagogically the minimal theory of Coulomb drag between two spatially-separated two-dimensional systems of massless Dirac fermions which are both away from the charge-neutrality point. This relies on second-order perturbation theory in the screened interlayer interaction and on Boltzmann transport theory. 
In this theoretical framework and in the low-temperature limit, we demonstrate that, to leading ({\it i.e.} quadratic) order in temperature, the drag transresistivity is completely insensitive to the precise intralayer momentum-relaxation mechanism ({\it i.e.} to the functional dependence of the scattering time on energy). We also provide analytical results for the low-temperature drag transresistivity for both cases of ``thick" and ``thin" spacers and for arbitrary values of the dielectric constants of the media surrounding the two Dirac-fermion layers. Finally, we present numerical results for the low-temperature drag transresistivity in the case in which one of the media surrounding the Dirac-fermion layers has a frequency-dependent dielectric constant. We conclude by suggesting an experiment that can potentially allow for the observation of departures from the canonical Fermi-liquid quadratic-in-temperature behavior of the transresistivity.
\end{abstract}
\maketitle

\section{Introduction}

Electron-electron interactions are a source of coupling  between closely spaced nano-electronic circuits.  This coupling has commanded a great deal of attention during the past thirty years or so, since it constitutes a potential alternative to the inductive and capacitive couplings of conventional electronics. Early on it was realized~\cite{pogrebinskii_1977, price_physicaB_1983} that ``Coulomb mutual scattering" between spatially separated electronic systems provides a mechanism to relax momentum that tends to equalize drift velocities. This {\it intrinsic} friction due to electron-electron interactions is modernly referred to as ``Coulomb drag"~\cite{jauho_prb_1993,zheng_prb_1993,kamenev_prb_1995, flensberg_prb_1995,badalyan_prb_2007,asgari_prb_2008,rojo_jpcm_1999}. Early experimental work was carried out by Gramila {\it et al.}~\cite{gramila_prl_1991} and by Sivan, Solomon, and Shtrikman~\cite{sivan_prl_1992} in semiconductor double quantum wells. 

In these experiments a constant current is imposed on the two-dimensional (2D) electron gas in one of the wells (the ``active" or ``drive" layer). If no current is allowed to flow in the other well (the ``passive" layer),  an electric field develops whose associated force cancels the frictional drag force exerted by the electrons in the active layer on the electrons in the passive one. The transresistance $\rho_{\rm D}$, defined as the ratio of the induced voltage in the passive layer to the applied current in the drive layer, directly measures the rate at which momentum is transferred from the current-carrying 2D electron gas to its neighbor. Coulomb drag is ultimately caused by {\it fluctuations} in the density of electrons in each layer since two-dimensional layers with uniformly distributed charge will not exert any frictional forces upon each other~\cite{zheng_prb_1993}.

The study of Coulomb-coupled 2D systems has now been revitalized
by advances which have made it possible to prepare robust and ambipolar 2D electron systems (ESs),
based on graphene~\cite{graphenereviews} layers
or on the surface states of topological insulators (TIs)~\cite{TIreviews},
that are described by an ultrarelativistic wave equation instead of the non-relativistic Schr\"{o}dinger equation.  

Single- and few-layer graphene systems can be produced, for example, by mechanical exfoliation
of thin graphite~\cite{tape} or by thermal decomposition of silicon carbide~\cite{SiCreviews}. 
Isolated graphene layers host massless-Dirac two-dimensional electron systems 
(MD2DESs) with a four-fold  (spin $\times$ valley)  flavor degeneracy,
whereas topologically-protected MD2DESs that have no additional spin or valley flavor
labels appear automatically~\cite{TIreviews} at the top and bottom surfaces
of a three-dimensional (3D) TI thin film.  
The protected surface states of 3D TIs are associated with spin-orbit interaction driven bulk band inversions.  3D TIs in a slab geometry offer two surface states that can be far enough
apart to make single-electron tunneling negligible, but close enough for Coulomb interactions between surfaces to be important.
Unhybridized MD2DES pairs can be realized in graphene by separating two 
layers by a dielectric~\cite{kim_prb_2011} (such as ${\rm Al}_2{\rm O}_3$) 
or by a few layers of a one-atom-thick insulator such as BN~\cite{dean_naturenano_2010,ponomarenko_naturephys_2011,britnell_science_2012,britnell_arXiv_2012}. 
In both cases interlayer hybridization is negligible and the nearby graphene layers are, from the point of view of single-particle physics, isolated.  Isolated graphene layers can be also found on the surface of bulk graphite~\cite{grapheneongraphite,li_natphys_2009} and in ``folded graphene"~\cite{schmidt_prb_2010}  (a natural byproduct of micromechanical exfoliation), or prepared by chemical vapor deposition~\cite{li_natphys_2009}. 
We use the term {\it double-layer graphene} (DLG) to
refer to a system with two graphene layers that are coupled only by Coulomb 
interactions, avoiding the term {\it bilayer graphene} which 
typically refers to two adjacent graphene layers in the crystalline Bernal-stacking configuration~\cite{graphenereviews}. 

DLG and TI thin films are both described at low energies by a Hamiltonian 
with two MD2DESs~\cite{graphenereviews} coupled only by Coulomb interactions, in the absence of single-particle tunneling. 
Coulomb drag between two spatially-separated MD2DESs has recently attracted a great deal of theoretical interest~\cite{tse_prb_2007,narozhny_prb_2007,katsnelson_prb_2011,peres_epl_2011,hwang_prb_2011,narozhny_arXiv_2011}. The calculations in Refs.~\onlinecite{tse_prb_2007,narozhny_prb_2007,katsnelson_prb_2011,peres_epl_2011,hwang_prb_2011,narozhny_arXiv_2011} refer to the regime in which {\it both} layers are either electron- or hole-doped. The Coulomb drag transresistivity in this case is negative and vanishes like $T^2$ at low temperatures. Despite the considerable amount of work published on the subject recently~\cite{tse_prb_2007,narozhny_prb_2007,katsnelson_prb_2011,peres_epl_2011,hwang_prb_2011,narozhny_arXiv_2011}, no clear consensus exists on the dependence of the drag transresistivity in the Fermi-liquid regime on carrier densities in the two layers, on the interlayer distance, and on the dielectric constants of the media surrounding the two layers. The main analytical results obtained earlier by other authors will be summarized below in Sect.~\ref{sect:summary}.

The Coulomb drag transresistivity between two MD2DESs in the regime in which one layer is electron doped and the other is hole doped has been recently calculated by Mink {\it et al.}~\cite{mink_arXiv_2011}. In this intriguing regime, the authors of Ref.~\onlinecite{mink_arXiv_2011} have found that $\rho_{\rm D}$ grows {\it logarithmically} upon lowering the temperature $T$ towards the critical temperature $T_{\rm c}$ for exciton condensation~\cite{Tc} (condensation of electron-hole pairs in a dipolar condensate).

Coulomb drag between two graphene sheets has been recently measured by Kim {\it et al.}~\cite{kim_prb_2011}. This first experimental study represents an important milestone since the authors of this work have shown that the ``strong-coupling" regime, {\it i.e.} the regime in which the interlayer distance $d$ is much smaller that the typical separation between two electrons in each layer, is easy to achieve experimentally with two, independently-contacted, graphene sheets~\cite{kim_prb_2011}. This study has indeed fueled the recent theoretical investigations of Coulomb drag between two MD2DESs mentioned above.

In this Article we present in a pedagogical fashion the minimal theory of Coulomb drag between two spatially-separated MD2DESs in the regime in which both layers are either electron- or hole-doped. We will be only concerned with the so-called ``Fermi-liquid regime" in which both layers are away from the charge neutrality point. Our theory relies on second-order perturbation theory in the screened interlayer interaction and on Boltzmann transport theory. In this theoretical framework and in the low-temperature limit, we demonstrate that, to leading ({\it i.e.} quadratic) order in temperature, the drag transresistivity is completely insensitive to the precise intralayer momentum-relaxation mechanism ({\it i.e.} to the functional dependence of the scattering time on energy). This is in contradiction with the findings reported in Refs.~\onlinecite{peres_epl_2011} and~\onlinecite{hwang_prb_2011}. We also provide new analytical results for the low-temperature drag transresistivity in both cases of ``thick" and ``thin" spacers, correcting in the latter case a mistake contained in Ref.~\onlinecite{katsnelson_prb_2011}. At odds with all the previous literature, our results hold true for arbitrary values of the dielectric constants of the media surrounding the two Dirac-fermion layers. Finally, we present numerical results for the low-temperature drag transresistivity in the case in which one of the media surrounding the two MD2DEs has a strongly-frequency-dependent dielectric constant. We conclude by suggesting an experiment with a DLG deposited on ${\rm SrTiO}_3$~\cite{couto_prl_2011} that can pave the way for the observation of departures from the canonical Fermi-liquid quadratic-in-temperature behavior of the transresistivity.

Our manuscript is organized as follows. In Sect.~\ref{sect:summary} we report a summary of the main analytical results which have been obtained by other authors. In Sect.~\ref{sect:theory} we present the model Hamiltonian and the most important basic definitions. In Sect.~\ref{sect:second_order} we present the Kubo formalism approach to the calculation of the drag conductivity, while in Sect.~\ref{sect:boltzmann} we present a series of simplifications that lead to the Boltmann-transport expression for the drag conductivity and resistivity. These two Sections do not contain original results but make the paper completely self-contained. Experts can skip Sects.~\ref{sect:second_order}-\ref{sect:boltzmann} and go directly to Sects.~\ref{sect:lowT}-\ref{sect:epsilon3omega}, which contain the most important results of this work and all our original results. In Sect.~\ref{sect:conclusions} we summarize our main findings and draw our main conclusions.

\begin{figure}[t]
\centering
\includegraphics[width=0.5\linewidth]{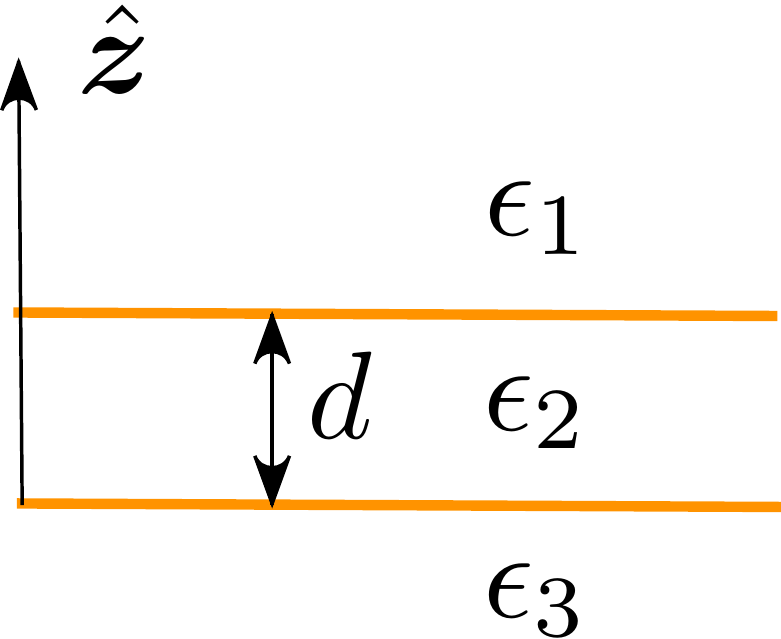}
\caption{(Color online) A side view of the double-layer system, 
which explicitly indicates the dielectric model used in these calculations. The two layers hosting massless Dirac fermions are located at $z=0$ and $z=d$. In a Coulomb-drag transport setup a constant current flow is imposed in one layer (the bottom one, say).  If no current is allowed to flow in the other layer, an electric field develops whose associated force cancels the frictional drag force exerted by the electrons in the bottom layer on the electrons in the top one.\label{fig:one}}
\end{figure}
\section{Summary of the main analytical results obtained earlier by other authors}
\label{sect:summary}

Despite the large body of theoretical work dedicated to Coulomb drag between two unhybridized MD2DESs~\cite{tse_prb_2007,narozhny_prb_2007,katsnelson_prb_2011,peres_epl_2011,hwang_prb_2011,narozhny_arXiv_2011}, no consensus appear to exist among different authors. In what follows we summarize the main analytical results that can be found in the existing literature.

1) Tse {\it et al.}~\cite{tse_prb_2007} studied Coulomb drag between two unhybrydized graphene sheets separated by a dielectric by employing Boltzmann transport theory. They neglected the spatial dependence of the dielectric constant in the ${\hat {\bm z}}$ direction (see Fig.~\ref{fig:one}) and assumed a momentum-independent scattering time. In the weak-coupling limit, {\it i.e.} in the limit in which the interlayer distance $d$ is much larger than the average distance between two electrons in each layer, Tse {\it et al.}~\cite{tse_prb_2007} demonstrated that the low-temperature drag resistivity is given by
\ber\label{eq:tse_2007_complete}
\rho_{\rm D}  &\to& - \frac{h}{e^2}\frac{\pi \zeta(3)}{32}~\frac{(k_{\rm B} T)^2}{\varepsilon_{{\rm F}, 1}\varepsilon_{{\rm F}, 2}}~\frac{1}{(q_{{\rm TF}, 1}d)(q_{{\rm TF}, 2}d)}\nonumber \\
&\times& \frac{1}{(k_{{\rm F}, 1}d)(k_{{\rm F},2}d)}~.
\eer
In Eq.~(\ref{eq:tse_2007_complete}) $q_{{\rm TF}, \ell}$ is the Thomas-Fermi screening wave vector, which is proportional to $k_{{\rm F}, \ell}$, the Fermi wave number in each layer. In the symmetric $n_1 = n_2$ case the previous equation yields
\be\label{eq:tse_2007_symmetric}
\rho_{\rm D} \propto - \frac{h}{e^2}~\frac{T^2}{n^3 d^4}~.
\ee

2) We now move on to summarize the main results by Peres {\it et al.}~\cite{peres_epl_2011}. The authors of this work used Boltzmann transport theory, took into account the momentum dependence of the scattering time and also the spatial dependence of the dielectric constant in the ${\hat {\bm z}}$ direction. While the main approach followed by these authors is numerical, they do also provide an analytical expression for the Coulomb drag transresistivity in the weak-coupling regime. For an intralayer scattering time that depends {\it linearly} on momentum, Peres {\it et al.}~\cite{peres_epl_2011} found
\be\label{eq:peres_2011}
\rho_{\rm D} \propto - \frac{h}{e^2}~\frac{T^2}{n^4 d^6},~{\rm for}~k_{\rm F} d \gg 1~.
\ee

3) Katsnelson~\cite{katsnelson_prb_2011} used Boltzmann transport theory, took into account, at least partially, the spatial dependence of the dielectric constant in the ${\hat {\bm z}}$ direction, but did not take into account the momentum dependence of the scattering time. The main results of Ref.~\onlinecite{katsnelson_prb_2011} are:
\be\label{eq:katsnelson_2011}
\left\{
\begin{array}{l}
{\displaystyle \rho_{\rm D} \propto - \frac{h}{e^2}~T^2~\frac{|\ln(nd^2)|}{n},~{\rm for}~k_{\rm F} d \ll 1}\vspace{0.2 cm}\\
{\displaystyle \rho_{\rm D} \propto - \frac{h}{e^2}~\frac{T^2}{n^3 d^4},~{\rm for}~k_{\rm F} d \gg 1}
\end{array}
\right.~.
\ee
Note that the weak-coupling result by Katsnelson is in agreement with that of Tse {\it et al.}~\cite{tse_prb_2007}.

4) Hwang {\it et al.}~\cite{hwang_prb_2011} used Boltzmann transport theory, took into account the momentum dependence of the scattering time but neglected the spatial dependence of the dielectric constant in the ${\hat {\bm z}}$ direction. For a {\it momentum-independent} intralayer scattering time they find:
\be\label{eq:Hwang_with_constant_scatteringtime}
\left\{
\begin{array}{l}
{\displaystyle \rho_{\rm D} \propto - \frac{h}{e^2} \frac{T^2}{n^2 d^2},~{\rm for}~k_{\rm F}d \ll 1}\vspace{0.2 cm} \\
{\displaystyle \rho_{\rm D} \propto - \frac{h}{e^2} \frac{T^2}{n^4 d^6},~{\rm for}~k_{\rm F}d \gg 1}
\end{array}
\right.~.
\ee
Note that the weak-coupling result in Eq.~(\ref{eq:Hwang_with_constant_scatteringtime}) differs from the result 
by Tse {\it et al.}~\cite{tse_prb_2007}. 

For an intralayer scattering time that depends {\it linearly} on momentum, Hwang {\it et al.}~\cite{hwang_prb_2011} find instead
\be\label{eq:Hwang_with_linear_scatteringtime}
\left\{
\begin{array}{l}
{\displaystyle \rho_{\rm D} \propto - \frac{h}{e^2} T^2\frac{|\ln(n d^2)|}{n},~{\rm for}~k_{\rm F}d \ll 1}\vspace{0.2 cm}\\
{\displaystyle \rho_{\rm D} \propto - \frac{h}{e^2}\frac{T^2}{n^3 d^4},~{\rm for}~k_{\rm F}d \gg 1}
\end{array}
\right.~.
\ee
By comparing Eq.~(\ref{eq:Hwang_with_constant_scatteringtime}) with Eq.~(\ref{eq:Hwang_with_linear_scatteringtime}), Hwang {\it et al.}~\cite{hwang_prb_2011} concluded that the functional dependence of $\rho_{\rm D}$ on $n$ and $d$ is very sensitive to the functional dependence of the intralayer scattering time on momentum.

5) Narozhny {\it et al.}~\cite{narozhny_arXiv_2011} have recently presented a systematic study of Coulomb drag between two MD2DESs which is based on perturbation theory in the dimensionless coupling constant $\alpha_{\rm ee} = e^2/(\hbar v)$, $v$ being the Dirac velocity. The authors of Ref.~\onlinecite{narozhny_arXiv_2011} did not consider the spatial dependence of the dielectric constant along the ${\hat {\bm z}}$ direction and discussed mostly the case of ``low doping" ({\it i.e.} the regime in which the chemical potential is comparable with or smaller than $k_{\rm B} T$). Since the focus of our Article is on the opposite regime, {\it i.e.} the Fermi-liquid regime, we now provide a short summary of the results of Narozhny {\it et al.}~\cite{narozhny_arXiv_2011} in this regime only. The authors of Ref.~\onlinecite{narozhny_arXiv_2011} have demonstrated that the energy dependence of the relaxation time is completely irrelevant in relation with the low-temperature drag transresistivity. In the weak-coupling limit Narozhny {\it et al.}~\cite{narozhny_arXiv_2011} found for the drag transresistivity the same doping- and interlayer separation-dependence as in Refs.~\onlinecite{tse_prb_2007} and~\onlinecite{katsnelson_prb_2011}. As density decreases, Narozhny {\it et al.}~\cite{narozhny_arXiv_2011} found that the drag coefficient acquires logarithmic corrections -- see Fig.~3 in Ref.~\onlinecite{narozhny_arXiv_2011}. In particular, in the limit $k_{\rm F} d \ll 1$, Eq.~(41) in Ref.~\onlinecite{narozhny_arXiv_2011} reads
\be\label{eq:narozhny}
\rho_{\rm D} \propto -\frac{h}{e^2}\frac{T^2}{n}\ln{\left(\frac{1}{N_{\rm f}\alpha_{\rm ee}}\right)}~.
\ee
As we will see below, our results agree with those of Narozhny {\it et al.}~\cite{narozhny_arXiv_2011}, and represent their generalization to the case of a 
finite Thomas-Fermi screening length ($\propto N_{\rm f}\alpha_{\rm ee}$) and spatially-dependent dielectric constants along the ${\hat {\bm z}}$ direction.

\section{Model Hamiltonian and basic definitions}
\label{sect:theory}

We consider two unhybridized layers of massless Dirac fermions, each one described at the noninteracting level by the following single-channel Hamiltonian ($\hbar =1$):
\begin{equation}\label{eq:MDFhamiltonian}
{\hat {\cal H}}_{\ell} =  v \sum_{\kv, \alpha, \beta} {\hat \psi}^\dagger_{\kv, \alpha, \ell} \left(\sigmav_{\alpha\beta} 
\cdot \kv\right) {\hat \psi}_{\kv, \beta, \ell}~.
\end{equation}
Here $v$ is the bare electron velocity, $\kv$ is the $\kv \cdot \pv$ momentum, $\alpha,\beta$ are (sublattice) pseudospin labels,
and $\sigmav_{\alpha\beta} = (\sigma^x_{\alpha\beta}, \sigma^y_{\alpha\beta})$ is a vector of Pauli matrices which act on the sublattice pseudospin degree-of-freedom. The field operator ${\hat \psi}^\dagger_{\kv, \alpha, \ell}$ (${\hat \psi}_{\kv, \alpha, \ell}$) creates (destroys) an electron with momentum $\kv$, pseudospin $\alpha$, and layer index $\ell =1,2$.

In each layer, the Hamiltonian (\ref{eq:MDFhamiltonian}) can be easily diagonalized by the matrix
\begin{equation}
{\cal U}({\bm k}) = \frac{1}{\sqrt{2}}
\left(
\begin{array}{cc}
e^{-i\varphi_{\bm k}/2} & e^{-i\varphi_{\bm k}/2}\\
e^{i\varphi_{\bm k}/2} & - e^{i\varphi_{\bm k}/2}
\end{array}
\right)~,
\end{equation}
where $\varphi_{\bm k}$ is the polar angle of the vector ${\bm k}$.

The $i$-th Cartesian component of the current-density operator ($i = x,y$) in the $\ell$-th layer is given by
\begin{equation}\label{eq:currentoperator}
{\hat J}^{i}_{{\bm q}, \ell} = \sum_{{\bm k}, \alpha, \beta} 
{\hat \psi}^\dagger_{{\bm k}-{\bm q}, \alpha, \ell}
(v \sigma^i_{\alpha\beta})
{\hat \psi}_{{\bm k}, \beta, \ell} ~.
\end{equation}
The Fermi wave number in the $\ell$-th layer is defined by
\be\label{eq:density}
k_{{\rm F}, \ell} \equiv \sqrt{\frac{4\pi n_\ell}{N_{{\rm f}, \ell}}}~,
\ee
where $n_\ell>0$ is the excess electron density in the $\ell$-th layer~\cite{phsymmetry} and $N_{{\rm f}, \ell}$ is a degeneracy factor ($N_{{\rm f}, \ell} =4$ for a graphene layer, accounting for spin and valley degeneracies, while $N_{{\rm f}, \ell} =1$ for a TI surface state). 

In the absence of disorder, the Green's function corresponding to the Hamiltonian in Eq.~(\ref{eq:MDFhamiltonian}) in the imaginary frequency axis is given by the following $2 \times 2$ matrix
\be \label{eq:green_funct}
G_\ell(\kv , i \omega) = \frac{(\mu_\ell + i \omega) \openone_\sigma + v \kv \cdot {\bm \sigma}}{(\mu_\ell + i\omega)^2 - v^2 \kv^2}~,
\ee
where $\openone_\sigma$ is the $2\times 2$ identity matrix in the sublattice-pseudospin representation and $\mu_\ell$ is the chemical potential in the $\ell$-th layer. In the zero-temperature limit 
$\mu_\ell \to \varepsilon_{{\rm F}, \ell} = v k_{{\rm F}, \ell}$, where $\varepsilon_{{\rm F}, \ell}$ is the Fermi energy in the $\ell$-th layer.

The disorder-free Green's function in the eigenstate representation reads instead
\be\label{eq:green_funct_diag}
{\cal G}_{\ell, \lambda}(\kv, i\omega) = \frac{1}{i\omega + \mu_\ell - \varepsilon_{{\bm k}, \lambda}}~,
\ee
where $\varepsilon_{{\bm k}, \lambda} = \lambda v |{\bm k}|$ are Dirac-band energies.

In what follows we will need the following matrix elements:
\ber \label{eq:rho_vertex}
\rho_{\kv\lambda, \kvp\lambda'} &\equiv& \big[ {\cal U}^\dagger(\kv) {\cal U}(\kvp)\big]_{\lambda\lambda'} \nonumber \\
&=& \frac{e^{i(\varphi_\kv -\varphi_{\kvp})/2} + \lambda \lambda' e^{-i(\varphi_\kv -\varphi_{\kvp})/2}}{2} ~,
\eer
and
\ber\label{eq:sigma_vertex}
\sigma^x_{\kv\lambda, \kvp\lambda'} &\equiv& \big[ {\cal U}^\dagger(\kv) \sigma^x {\cal U}(\kvp)\big]_{\lambda\lambda'} \nonumber\\
&=& \frac{\lambda  e^{-i(\varphi_\kv +\varphi_{\kvp})/2} + \lambda' e^{i(\varphi_\kv +\varphi_{\kvp})/2}}{2}~.
\eer

The two MD2DESs described by Eq.~(\ref{eq:MDFhamiltonian}) 
are coupled electrostatically by long-range Coulomb interactions, which are influenced by the layered dielectric environment (see Fig.~\ref{fig:one}). The coupling Hamiltonian reads
\be
{\hat {\cal H}}_{\rm ee} = \frac{1}{2 S}\sum_{{\bm q}, \ell \neq \ell'} V_{\ell \ell'}(q){\hat \rho}_{{\bm q}, \ell} {\hat \rho}_{-{\bm q}, \ell'}~,
\ee
where
\begin{equation}\label{eq:densityoperator}
{\hat \rho}_{{\bm q}, \ell} = \sum_{{\bm k}, \alpha} {\hat \psi}^\dagger_{{\bm k} - {\bm q}, \alpha, \ell}{\hat \psi}_{{\bm k}, \alpha, \ell}
\end{equation}
is the density-operator for the $\ell$-th layer and $V_{\ell \ell'}(q)$ (with $\ell \neq \ell'$) is the 2D Fourier transform of the interlayer Coulomb interaction
\be\label{eq:v12} 
V_{12}(q) = V_{21}(q) = \frac{8\pi e^2}{q D(q)}~\epsilon_2~.
\ee
Here
\begin{equation}\label{eq:denominator}
D(q) =  [(\epsilon_1 + \epsilon_2) (\epsilon_2 + \epsilon_3) e^{qd} + 
(\epsilon_1 - \epsilon_2) (\epsilon_2 - \epsilon_3) e^{-qd} ]~.
\end{equation}

For future purposes, we introduce the dynamically screened interlayer interaction $U_{12}(q,\omega)$, which, at the random phase approximation (RPA) level, is given by~\cite{zheng_prb_1994,Giuliani_and_Vignale}
\begin{equation}\label{eq:W12}
U_{12}(q, \omega) = \frac{V_{12}(q)}{\varepsilon(q,\omega)}~,
\end{equation}
where
\begin{eqnarray}\label{eq:dynamicalRPA}
\varepsilon(q,\omega) &=& [1-V_{11}(q) \chi^{(0)}_1(q,\omega)][1-V_{22}(q)\chi^{(0)}_2(q,\omega)] \nonumber\\
&-& V^2_{12}(q)\chi^{(0)}_1(q,\omega)\chi^{(0)}_2(q,\omega)
\end{eqnarray}
is the RPA dynamical dielectric function. 

In Eq.~(\ref{eq:dynamicalRPA}), $\chi^{(0)}_\ell(q,\omega)$ is the well-known~\cite{hwang_prb_2007,barlas_prl_2007,wunsch_njp_2006} density-density (Lindhard) response function of a noninteracting MD2DEs at arbitrary doping $n_\ell$. The Coulomb interaction in the $\ell =1$ (top) layer is given by
\begin{equation}\label{eq:v11}
V_{11}(q) = \frac{4\pi e^2}{q D(q)} [ (\epsilon_2 + \epsilon_3) e^{qd} + 
 (\epsilon_2 - \epsilon_3) e^{-qd}]~,
\end{equation}
while the Coulomb interaction in the bottom layer, $V_{22}(q)$, can be simply obtained from $V_{11}(q)$ 
by interchanging $\epsilon_3 \leftrightarrow \epsilon_1$.

Eqs.~(\ref{eq:v12}), (\ref{eq:denominator}), and~(\ref{eq:v11}) have first appeared in Ref.~\onlinecite{profumo_prb_2010} and their explicit derivation has been reported in 
Ref.~\onlinecite{katsnelson_prb_2011}. Notice that in the ``uniform" $\epsilon_1 = \epsilon_2 = \epsilon_3 \equiv \epsilon$ limit we recover the familiar expressions $V_{11}(q) = V_{22}(q) \to 2\pi e^2/(\epsilon q)$ and $V_{12}(q) = V_{21}(q) \to V_{11}(q)\exp(-qd)$. 
Most of the previous work on Coulomb drag in DLG has assumed this limit, which 
rarely applies experimentally. 

The aim of this Article is to present a theory of Coulomb drag, which is valid up to {\it second order} in the dynamically-screened interaction $U_{12}(q,\omega)$, for the system described by the Hamiltonian
\be\label{eq:hamtotal}
{\hat {\cal H}} = \sum_{\ell} {\hat {\cal H}}_\ell + {\hat {\cal H}}_{\rm ee}~.
\ee

Note that we are not including in Eq.~(\ref{eq:hamtotal}) any term describing {\it intralayer} electron-electron interactions. 
These can be treated in an approximate fashion by invoking Landau's theory of normal Fermi liquids~\cite{Giuliani_and_Vignale}, {\it i.e.} by renormalizing the microscopic parameters of the intralayer Hamiltonian ${\hat {\cal H}}_\ell$. For example, in Eq.~(\ref{eq:MDFhamiltonian}) 
one can use the renormalized quasiparticle velocity~\cite{velocityenhancement}, $v^\star_\ell$, instead of the bare velocity $v$. Anyway, a treatment of the impact of intralayer interactions on the Coulomb drag transresistivity is well beyond the scope of this Article.

\begin{figure}[t]
\centering
\includegraphics[width=0.9\linewidth]{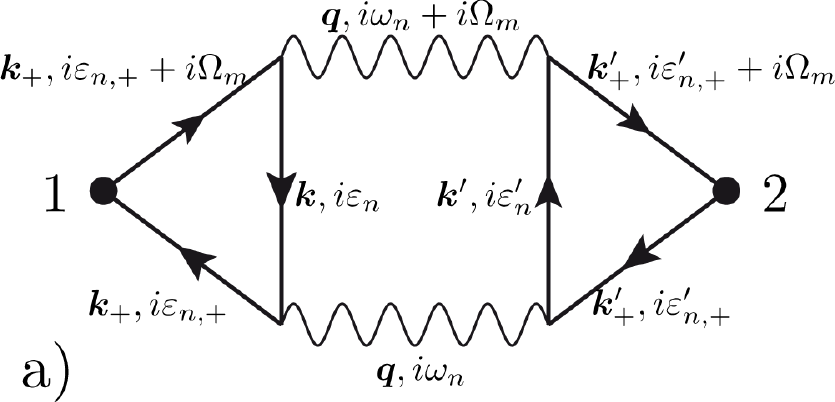}\vspace{0.6 cm}\\
\includegraphics[width=0.9\linewidth]{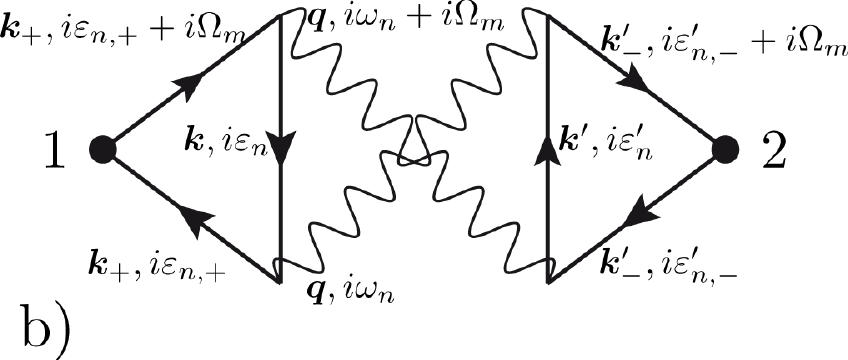}
\caption{Second-order Aslamazov-Larkin-type diagrams contributing to the Coulomb drag conductivity (finite-temperature Matsubara formalism). Solid lines denote the single-particle Green's function in the presence of disorder. Wavy lines denote the {\it screened} interlayer interaction $U_{12}$ in Eq.~(\ref{eq:W12}). Black dots in the triangular portions of the diagrams denote vertices of current operators in the two layers, which are both proportional to the Pauli matrix $\sigma^x$, say, if one is interested in the longitudinal drag conductivity. Finally, 
${\bm k}_\pm = {\bm k} \pm {\bm q}$, ${\bm k}'_\pm = {\bm k}' \pm {\bm q}$, $i \varepsilon_{n, \pm} = i \varepsilon_n \pm i \omega_n$, 
and $i \varepsilon'_{n, \pm} = i \varepsilon'_n \pm i \omega_n$. \label{fig:two}}
\end{figure}
\section{Kubo-formula approach to the calculation of the drag conductivity} 
\label{sect:second_order}

Coulomb drag starts at second order in a perturbative expansion for the ``drag conductivity" $\sigma_{\rm D}$ in powers of the interlayer Coulomb interaction. To avoid infrared pathologies (stemming from the long-range nature of the Coulomb interaction) of certain integrals that appear in the theory, Coulomb interactions must be screened: from now on we will work with the dynamically-screened interaction $U_{12}$ introduced in Eq.~(\ref{eq:W12}) rather than with the bare potential $V_{12}$ in Eq.~(\ref{eq:v12}). 

The two Aslamazov-Larkin-type diagrams~\cite{varlamovlarkinbook} that contribute to $\sigma_{\rm D}$ up to second order in $U_{12}$ are depicted in Fig.~\ref{fig:two}. In this figure, solid lines represent 
disordered propagators in the sublattice-pseudospin representation, while wavy lines represent the screened Coulomb interaction $U_{12}$.
Finally, the vertices (black dots in the triangular diagrams) are current operators, one in each layer, evaluated at ${\bm q} = {\bm 0}$ from the very beginning since we are interested in the drag conductivity in the uniform limit. Both current operators can be taken along the ${\hat {\bm x}}$ direction since we are interested in the current response of the ``passive" layer in the ${\hat {\bm x}}$ direction to an electric field applied in the same direction in the ``active" layer, {\it i.e.} we are interested in the {\it longitudinal} drag conductivity.

Straightforward algebraic manipulations lead to the following compact expression for the sum of the two diagrams in Fig.~\ref{fig:two}:
\begin{widetext}
\be \label{eq:sigma_second}
\sigma_{\rm D}(i \Omega_m) =\frac{e^2}{2 \Omega_m} \int\frac{d^2{\bm q}}{(2\pi)^2} 
\frac{1}{\beta} \sum_{\omega_n} U_{12}(q, i\omega_n) U_{12}(q, i\omega_n+i\Omega_m)
\Gamma_1({\bm q},i\omega_n+i\Omega_m,i\omega_n) \Gamma_2({\bm q},i\omega_n,i\omega_n+i\Omega_m)
~,
\ee
\end{widetext}
where $\beta= (k_{\rm B} T)^{-1}$ is the usual thermal factor, $\Omega_m$ and $\omega_n$ are {\it bosonic} Matsubara frequencies and
\begin{widetext}
\ber \label{eq:Gamma_second_1}
\Gamma_\ell({\bm q}, i\omega_1, i\omega_2) &=& \frac{v N_{{\rm f}, \ell}}{\beta} \sum_{\varepsilon_n} \int\frac{d^2{\bm k}}{(2\pi)^2}
{\rm Tr}\big[ G_\ell({\bm k},i\varepsilon_n) G_\ell({\bm k} + {\bm q},i\varepsilon_n+i\omega_2) \sigma^x 
G_\ell({\bm k} + {\bm q},i\varepsilon_n+i\omega_1) 
\nonumber\\
&+& G_\ell({\bm k}, i\varepsilon_n) G_\ell({\bm k} - {\bm q},i\varepsilon_n - i\omega_1) \sigma^x G_\ell({\bm k} - {\bm q},i\varepsilon_n-i\omega_2) \big]
\eer
\end{widetext}
is the so-called ``non-linear susceptibility". Here $\varepsilon_n$ is a {\it fermionic} Matsubara frequency, 
the index $\ell =1,2$ refers to the layer degree of freedom, and the symbol ${\rm Tr}[\dots]$ denotes a trace over sublattice-pseudospin indices.

\begin{figure}
\centering
\includegraphics[width=0.8\linewidth]{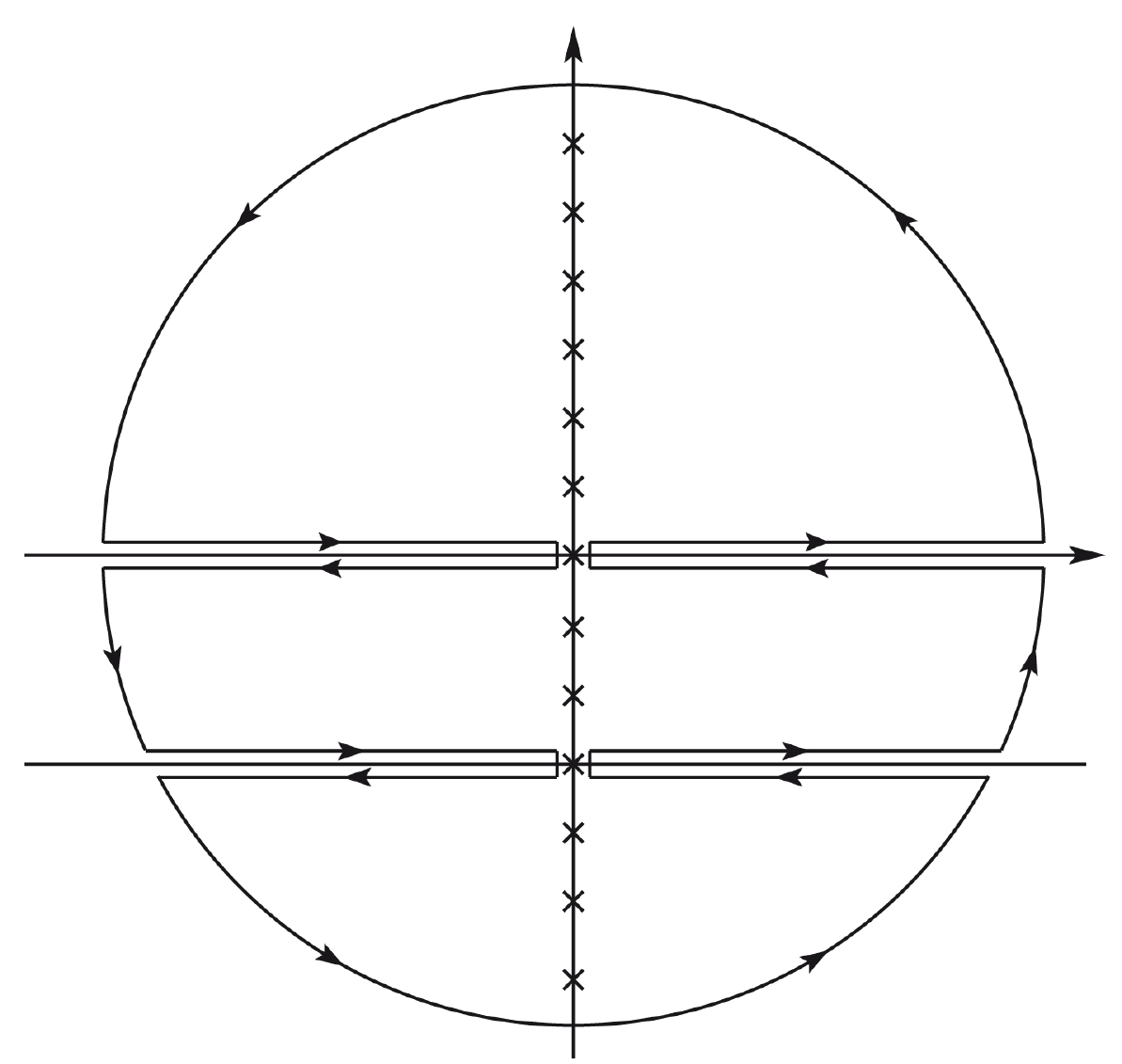}
\caption{The contour of integration ${\cal C}$ chosen to carry out the Matsubara sum in Eq.~(\ref{eq:J_second_def}). Note the branch cuts on the real axis and in the lower half of the complex plane at 
$\Im m(z) = - i\Omega_m$. The crosses denote the poles of the Bose-Einstein thermal factor $n_{\rm B}(z)$.\label{fig:three}}
\end{figure}

We now need to perform the analytic continuation $i\Omega_m \to \Omega + i 0^+$ in Eq.~(\ref{eq:sigma_second}) and then take the d.c. 
$\Omega \to 0$ limit. It is very well known that the analytic continuation must be carried out {\it after} performing the sum over the 
Matsubara frequencies $\omega_n$. Let us assume that $U_{12}(q, z)$, seen as a function of the complex frequency $z$, is {\it analytic}, {\it i.e.} that it has {\it no} branch cuts. Under this assumption, the function $\Gamma_\ell({\bm q}, z, z')$ has {\it two} branch cuts when $z$ or $z'$ are on the real axis and is analytic elsewhere, including the points $z=0$ and $z'=0$. We then introduce
\be
f({\bm q}, z, z')= U_{12}(q,z) U_{12}(q, z') \Gamma_1({\bm q},z',z) \Gamma_2({\bm q},z,z')
~,
\ee
and perform the Matsubara sum in Eq.~(\ref{eq:sigma_second}) using a standard procedure:
\ber \label{eq:J_second_def}
{\cal J}(i\Omega_m) &\equiv& \frac{1}{\beta} \sum_{\omega_n} f({\bm q} , i\omega_n, i\Omega_m + i\omega_n)  \nonumber \\
&=& \oint_{\cal C} \frac{dz}{2\pi i} n_{\rm B}(z) f({\bm q}, z, i\Omega_m + z)~,
\eer
where $n_{\rm B}(z)=(e^{\beta z}-1)^{-1}$ is the Bose-Einstein occupation factor on the complex plane and ${\cal C}$ is an appropriate contour which excludes the branch cuts of the integrand (see Fig.~\ref{fig:three}), located at $\Im m(z)=0, -i\Omega_m$. Notice that the contour ${\cal C}$ as defined in Fig.~\ref{fig:three} {\it includes} the points $z=0$ and $z=-i\Omega_m$, where the integrand is analytic.  
We obtain
\begin{widetext}
\ber
{\cal J}(i\Omega_m) &=& {\cal P}\int_{-\infty}^{+\infty} \frac{d\omega}{2\pi i} n_{\rm B}(\omega)\big[
f({\bm q},\omega^+, i\Omega_m+\omega) - f(i\Omega_m,\omega^-, i\Omega_m+\omega)
\nonumber\\
&+& f(i\Omega_m, \omega - i\Omega_m, \omega^+) - f(i\Omega_m, \omega - i\Omega_m, \omega^-)
\big]
~,
\eer
\end{widetext}
since only the integrals around the branch cuts contribute to ${\cal J}(i\Omega_m)$, while the integral over the circle vanishes when its radius is sent to infinity. Here $\omega^\pm=\omega \pm i 0^+$ and ${\cal P}$ stands for the Cauchy principal value, {\it i.e.} the point $\omega=0$ is excluded from the integration.

We are now ready to perform the analytical continuation to real frequencies. After some algebra we find
\begin{widetext}
\ber \label{eq:J_final}
{\cal J}(\Omega + i \epsilon) &=& {\cal P}\int_{-\infty}^{+\infty} \frac{d\omega}{2\pi i} \Big\{
\big[n_{\rm B}(\Omega+\omega) - n_{\rm B}(\omega)\big] f({\bm q},\omega^-, \Omega+\omega^+)
\nonumber\\
&+& n_{\rm B}(\omega) f({\bm q}, \omega^+, \Omega+\omega^+)
 - n_{\rm B}(\Omega+\omega) f({\bm q},\omega^-, \Omega+\omega^-)
\Big\}
~,
\eer
\end{widetext}

We finally take the d.c. limit $\Omega\to 0$ in Eq.~(\ref{eq:J_final}) and we obtain
\be
\lim_{\Omega\to 0}{\cal J}(\Omega + i \epsilon) \to \Omega~ \int_{-\infty}^{+\infty} \frac{d\omega}{2\pi i} 
\frac{\partial n_{\rm B}(\omega)}{\partial \omega} f({\bm q},\omega^-, \omega^+)
~,
\ee
since the terms in the last line of Eq.~(\ref{eq:J_final}) vanish at least like $\Omega^2$. Thus the drag conductivity in the d.c. limit 
reads~\cite{kamenev_prb_1995}
\begin{widetext}
\be\label{eq:conductivity_end}
\sigma_{\rm D} = \frac{\beta e^2}{16\pi} \int\frac{d^2{\bm q}}{(2\pi)^2}
\int_{-\infty}^{+\infty} d\omega \frac{|U_{12}(q,\omega)|^2}{\sinh^2(\beta\omega/2)}
\Gamma_1({\bm q},\omega^+,\omega^-) \Gamma_2({\bm q},\omega^-,\omega^+)~.
\ee
\end{widetext}
This is the central result of this Section and was first obtained in the context of Coulomb drag between ordinary 2D parabolic-band electron gases -- see {\it e.g.} Ref.~\onlinecite{kamenev_prb_1995}.

\subsection{The non-linear susceptibility} \label{sect:nls}
Let us go back to the definition of the non-linear susceptibility $\Gamma_\ell$ given in Eq.~(\ref{eq:Gamma_second_1}). We introduce the following definition
\ber \label{eq:g_def}
g_\ell({\bm q}, i\varepsilon, i\varepsilon', i\varepsilon'') &=& v N_{{\rm f}, \ell}\int\frac{d^2{\bm k}}{(2\pi)^2}
{\rm Tr}\big[ G_\ell({\bm k},i\varepsilon)\nonumber\\
&\times& G_\ell({\bm k} + {\bm q},i\varepsilon')\sigma^x G_\ell({\bm k} + {\bm q},i\varepsilon'')\big]\nonumber\\
\eer
so that Eq.~(\ref{eq:Gamma_second_1}) reads
\ber \label{eq:Gamma_second_g}
\Gamma_\ell({\bm q}, i\omega_1, i\omega_2) &=& \frac{1}{\beta} \sum_{\varepsilon_n} \big[
g_\ell({\bm q}, i\varepsilon_n, i\varepsilon_n+i\omega_2, i\varepsilon_n+i\omega_1) \nonumber\\
&+& g_\ell({\bm q}, i\varepsilon_n, i\varepsilon_n-i\omega_1, i\varepsilon_n-i\omega_2)
\big] 
\nonumber\\
&\equiv& {\cal J}_g ({\bm q}, i \omega_1 , i \omega_2) \nonumber \\
&+& {\cal J}_g (-{\bm q}, -i \omega_2 , -i \omega_1)
~.
\eer
We here remind the reader that $\omega_1$ and $\omega_2$ ($\varepsilon_n$) are bosonic (fermionic) Matsubara frequencies. Let us now concentrate on the first term on the r.h.s. of Eq.~(\ref{eq:Gamma_second_g}), {\it i.e.} on ${\cal J}_g ({\bm q}, i \omega_1 , i \omega_2)$. 
For the sake of simplicity, in what follows we omit to indicate explicitly the ${\bm q}$-dependence of the functions 
$g_\ell({\bm q}, i\varepsilon, i\varepsilon', i\varepsilon'')$ and ${\cal J}_g ({\bm q}, i \omega_1 , i \omega_2)$ every time we write equations that involve them. The complete notation will be restored only at the end of the mathematical manipulations that we perform on Eq.~(\ref{eq:Gamma_second_g}).

We first perform the fermionic Matsubara sum in the first term on the r.h.s. of Eq.~(\ref{eq:Gamma_second_g}) by the standard procedure:
\ber \label{eq:J_g}
{\cal J}_g(i\omega_1, i\omega_2) &=& \frac{1}{\beta} \sum_{\varepsilon_n} g_\ell(i\varepsilon_n, i\varepsilon_n+i\omega_2, i\varepsilon_n+i\omega_1)
\nonumber\\
&=& \oint_{\cal C} \frac{d z}{2\pi i} n_{\rm F}(z) g_\ell(z, z+i\omega_2,z+i\omega_1)~,\nonumber\\
\eer
where $n_{\rm F}(z) = (e^{\beta z} + 1)^{-1}$ is the Fermi-Dirac occupation factor on the complex plane. The function $g(z, z+i\omega_2,z+i\omega_1)$ has three branch cuts in the complex plane located at $\Im m(z) = 0, -i\omega_1, -i\omega_2$. Choosing a suitable contour of integration ${\cal C}$, which encircles only the poles of $n_{\rm F}(z)$ and leaves outside the branch cuts (in complete analogy with the contour drawn in Fig.~\ref{fig:three}), we find
\ber
{\cal J}_g(i\omega_1, i\omega_2) &=& \int_{-\infty}^{+\infty} \frac{d\varepsilon}{2\pi i} n_{\rm F}(\varepsilon) 
\big[
g_\ell(\varepsilon^+, \varepsilon +i\omega_2,\varepsilon +i\omega_1) \nonumber\\
&-& g_\ell(\varepsilon^-, \varepsilon +i\omega_2,\varepsilon +i\omega_1)
\nonumber\\
&+&
g_\ell(\varepsilon - i\omega_1, \varepsilon^+ +i\omega_2 - i\omega_1,\varepsilon^+) \nonumber \\
&-& g_\ell(\varepsilon - i\omega_1, \varepsilon^- +i\omega_2 - i\omega_1,\varepsilon^-)
\nonumber\\
&+&
g_\ell(\varepsilon - i\omega_2, \varepsilon^+,\varepsilon^+ + i\omega_1 - i\omega_2) \nonumber \\
&-& g_\ell(\varepsilon - i\omega_2, \varepsilon^-,\varepsilon^- + i\omega_1 - i\omega_2)
\big]~.
\eer
Here $\varepsilon^\pm = \varepsilon \pm i0^+$ and we used that $n_{\rm F}(\varepsilon+i\omega_m) = n_{\rm F}(\varepsilon)$ if $\omega_m$ is a bosonic Matsubara frequency. Notice that we are allowed to remove the index $\pm$ from $\varepsilon^\pm$ when it is summed to $i\omega_1$ or $i\omega_2$ (since they will be analytically continued to $\omega^\pm$) but not when it is summed to $i\omega_1-i\omega_2$. Analytically continuing $i\omega_1 \to \omega_1^+$ and $i\omega_2 \to \omega_2^-$ [to obtain {\it e.g.} $\Gamma_1({\bm q}, \omega^+, \omega^-)$ in the integrand on the r.h.s. of Eq.~(\ref{eq:conductivity_end})] we finally obtain
\begin{widetext}
\ber \label{eq:J_g_real}
{\cal J}_g(\omega_1^+, \omega_2^-) &=& \int_{-\infty}^{+\infty} \frac{d\varepsilon}{2\pi i} n_{\rm F}(\varepsilon) 
\left[
g_\ell(\varepsilon^+, \varepsilon^- + \omega_2,\varepsilon^+ + \omega_1) - g_\ell(\varepsilon^-, \varepsilon^- + \omega_2,\varepsilon^+ + \omega_1)
\right.
\nonumber\\
&+&
g_\ell(\varepsilon^- - \omega_1, \varepsilon^- + \omega_2 - \omega_1,\varepsilon^+) - g_\ell(\varepsilon^- - \omega_1, \varepsilon^- + \omega_2 - \omega_1,\varepsilon^-)
\nonumber\\
&+&
\left.
g_\ell(\varepsilon^+ - \omega_2, \varepsilon^+,\varepsilon^+ + \omega_1 - \omega_2) - g_\ell(\varepsilon^+ - \omega_2, \varepsilon^-,\varepsilon^- + \omega_1 - \omega_2)
\right]
~.
\eer
\end{widetext}
We observe that the second term on the second line and the first term on the third line of Eq.~(\ref{eq:J_g_real}) are exactly zero since they involve products of three Green's functions with poles on the same half of the complex plane. After some algebraic manipulations we get
\begin{widetext}
\ber
{\cal J}_g(\omega_1^+, \omega_2^-) &=& \int_{-\infty}^{+\infty} \frac{d\varepsilon}{2\pi i}
\left\{
\left[ n_{\rm F}(\varepsilon) - n_{\rm F}(\varepsilon+\omega_2) \right] g_{\ell}(\varepsilon^+, \varepsilon^- +\omega_2, \varepsilon^++\omega_1)
\right.
\nonumber\\
&+&
\left.
\left[ n_{\rm F}(\varepsilon+\omega_1) - n_{\rm F}(\varepsilon) \right] g_{\ell}(\varepsilon^-, \varepsilon^- +\omega_2, 
\varepsilon^++\omega_1)
\right\}
~.
\eer
\end{widetext}
We now take the limit $\omega_1, \omega_2 \to \omega$ and recast Eq.~(\ref{eq:J_g}) in the following form:
\begin{widetext}
\ber\label{gam_final}
{\cal J}_g(\omega^+, \omega^-) &=& v N_{{\rm f}, \ell}\int_{-\infty}^{+\infty} \frac{d\varepsilon}{2\pi i}
\left[ n_{\rm F}(\varepsilon+\omega) - n_{\rm F}(\varepsilon) \right]\nonumber\\
&\times&\int\frac{d^2{\bm k}}{(2\pi)^2}
{\rm Tr}\Big\{ \big[G^{\rm A}_\ell({\bm k},\varepsilon) - G^{\rm R}_\ell({\bm k},\varepsilon)\big] G^{\rm A}_\ell({\bm k} + {\bm q},\varepsilon+\omega) \sigma^x 
G^{\rm R}_\ell({\bm k} + {\bm q},\varepsilon+\omega)
\Big\}
~,
\eer
\end{widetext}
where we have introduced the retarded (advanced) Green's function $G^{{\rm R}({\rm A})}_\ell({\bm k} , \varepsilon)$.

The second term on the right hand of Eq.~(\ref{eq:Gamma_second_g}) can be treated in an analogous manner. The final expression for the non-linear susceptibility reads
\begin{widetext}
\ber\label{eq:gamma_end}
\Gamma_\ell({\bm q} , \omega^+ , \omega^- ) &=& v N_{{\rm f}, \ell} \int_{-\infty}^{+\infty}\frac{d \varepsilon}{2\pi i } 
\big[n_{\rm F} (\varepsilon  +\omega) - n_{\rm F}(\varepsilon )\big] \int \frac{d^2 {\bm k}}{(2\pi)^2} {\rm Tr}\Big\{\big[G^{\rm A}_\ell({\bm k} , \varepsilon ) - G^{\rm R}_\ell({\bm k}, \varepsilon ) \big] G^{\rm A}_\ell({\bm k} + {\bm q} , \varepsilon + \omega ) \sigma^x 
\nonumber\\
&\times& G^{\rm R}_\ell({\bm k } + {\bm q} , \varepsilon + \omega ) \Big\} + \Big\{ ({\bm q} , \omega) \rightarrow (-{\bm q } , - \omega) \Big\}~.
\eer
\end{widetext}
Eq.~(\ref{eq:gamma_end}) together with Eq.~(\ref{eq:conductivity_end}) represent the most important results of this Section and are the starting point for the calculation of the drag transresistivity $\rho_{\rm D}$. 

\section{Boltzmann-transport limit} 
\label{sect:boltzmann}

Eqs.~(\ref{eq:conductivity_end}) and~(\ref{eq:gamma_end}) need to be simplified for any practical purpose and for a quantitative estimate of the Coulomb drag transresistivity.

We first switch from the off-diagonal sublattice-pseudospin representation of the Green's functions to the diagonal representation. In the latter representation Eq.~(\ref{eq:gamma_end}) reads
\begin{widetext}
\ber\label{eq:gamma_diagonal}
\Gamma_\ell ({\bm q},\omega^+, \omega^-) &=& vN_{{\rm f}, \ell}\int_{-\infty}^{+\infty} \frac{d \varepsilon}{2\pi i} \int \frac{d^2 {\bm k}}{(2\pi)^2} \sum_{\lambda , \lambda' , \lambda''} \Big\{\big[n_{\rm F}(\varepsilon + \omega ) - n_{\rm F}(\varepsilon )\big] \big[{\cal G}^{\rm A}_{\ell, \lambda} ({\bm k} , \varepsilon) - {\cal G}^{\rm R}_{\ell, \lambda}({\bm k} , \varepsilon ) \big]
\nonumber\\
&\times& {\cal G}^{\rm A}_{\ell, \lambda'}({\bm k} + {\bm q}, \varepsilon + \omega) 
{\cal G}^{\rm R}_{\ell, \lambda''}({\bm k} + {\bm q},\varepsilon + \omega)~\rho_{{\bm k} \lambda, {\bm k } +{\bm q} \lambda'} 
\sigma^x_{{\bm k} +{\bm q} \lambda' , {\bm k } + {\bm q} \lambda''}\rho_{{\bm k} +{\bm q} \lambda'' , {\bm k} \lambda}\Big\}
\nonumber\\
&+& \Big\{ ({\bm q}, \omega) \rightarrow (-{\bm q} , -\omega) \Big\}~,
\eer
\end{widetext}
where we used the definitions in Eqs.~(\ref{eq:rho_vertex})-(\ref{eq:sigma_vertex}) for the density and current vertices. 

We now show that the terms with $\lambda' \neq \lambda''$ do {\it not} contribute to $\Gamma_\ell ({\bm q},\omega^+, \omega^-)$. 

Let us indeed consider all the terms in the triple sum in Eq.~(\ref{eq:gamma_diagonal}) in which $\lambda'' = - \lambda' \equiv {\bar \lambda}'$. We first recall that $\sigma^x_{{\bm k} + {\bm q} \lambda', {\bm k} + {\bm q} {\bar \lambda}'} = i \lambda '  \sin(\varphi_{{\bm k} + {\bm q }})$ and 
\ber
\rho_{{\bm k} \lambda , {\bm k} + {\bm q } \lambda'} \rho_{{\bm k} + {\bm q} {\bar \lambda}' , {\bm k} \lambda}= \frac{i \lambda \lambda '}{2} \sin (\varphi_{{\bm k}} - \varphi_{{\bm k } + {\bm q }})
~.
\eer
We then consider the term on the third line of Eq.~(\ref{eq:gamma_diagonal}) and perform the following changes of variables: ${\bm k} \to - {\bm k}$ and $\varepsilon \to - \varepsilon $.
Using the fact that $\sin (\varphi_{-{\bm k} - {\bm q}}) = - \sin (\varphi_{{\bm k} + {\bm q}})$ and $\sin(\varphi_{-{\bm k}} - \varphi_{-{\bm k}-{\bm q}}) = \sin (\varphi_{{\bm k}} -\varphi_{{\bm k} + {\bm q}})$, we can write the off-diagonal (OD) contribution ($\lambda'' = -\lambda'$) to $\Gamma^{\rm OD}_\ell ({\bm q},\omega^+, \omega^-)$ as
\begin{widetext}
\ber
\Gamma^{\rm OD}_\ell ({\bm q},\omega^+, \omega^-) &=& v N_{{\rm f}, \ell}\int_{\infty}^{+\infty}\frac{d \varepsilon}{2\pi i} \int \frac{d^2 {\bm k}}{(2\pi)^2} \sum_{\lambda , \lambda' } \Big(- \frac{\lambda }{2}\Big)\Big\{\big[n_{{\rm F}}(\varepsilon + \omega )+ n_{{\rm F}} (- \varepsilon - \omega ) -  n_{{\rm F}}(\varepsilon )- n_{{\rm F}} (-\varepsilon)\big] \nonumber\\
&\times &\big[{\cal G}^{\rm A}_{\ell, \lambda}({\bm k} , \varepsilon) - {\cal G}^{\rm R} _{\ell, \lambda}({\bm k} , \varepsilon ) \big]
{\cal G}^{\rm A}_{\ell, \lambda'}({\bm k} + {\bm q} , \varepsilon + \omega) {\cal G}^{\rm R}_{\ell, {\bar \lambda}'} ( {\bm k} + {\bm q } , \varepsilon + \omega ) \sin (\varphi_{{\bm k}} - \varphi_{{\bm k} + {\bm q}}) \sin (\varphi_{{\bm k} + {\bm q}}) \Big\}~.
\eer
\end{widetext}
Since $n_{\rm F}(z) + n_{\rm F}(-z) = 1$, the term in square brackets is identically zero. The OD contribution to 
$\Gamma_\ell({\bm q},\omega^+, \omega^-)$ thus vanishes. From now on we can set $\lambda'' = \lambda'$ in Eq.~(\ref{eq:gamma_diagonal}) and sum only over $\lambda$ and $\lambda'$. 

We can further simplify Eq.~(\ref{eq:gamma_diagonal}) by assuming that, in the presence of weak disorder, the Green's function in the $\ell$-th layer can be well approximated by the expression,
\be\label{eq:finitelifetimeeffect}
{\cal G}^{{\rm R}({\rm A})}_{\ell, \lambda}({\bm k}, \omega) \approx \Bigg[\omega - \xi^{(\ell)}_{{\bm k}, \lambda} \pm \frac{i}{2 \tau_\ell({\bm k})}\Bigg]^{-1}~,
\ee
where $\tau_\ell({\bm k})$ represents a momentum-dependent scattering time in the $\ell$-th layer and $\xi^{(\ell)}_{{\bm k}, \lambda} \equiv \lambda v |{\bm k}|- \mu_\ell$ are 
Dirac-band energies measured from the chemical potential $\mu_\ell$ of the $\ell$-th layer.

Note that here $\tau_\ell({\bm k})$ should not be interpreted as a one-particle scattering time, but rather as the {\it transport scattering time}~\cite{flensberg_prb_1995}. This statement can be justified within the Kubo formalism by including disorder-related vertex corrections, or, in a much more transparent way, by using the Boltzmann transport equation, see Appendix~\ref{appendix:boltzmann}.

Below, we will assume that the transport scattering time $\tau_\ell({\bm k})$ is isotropic, {\it i.e.} that it depends only on $k = |{\bm k}|$, but allow different scattering times in the two layers. The specific functional dependence of $\tau_\ell$ on $k$ depends on a particular model of intralayer impurity scattering. In the simple case of a momentum-independent scattering time we recover the usual ``relaxation time approximation".  A scattering time which depends linearly on momentum, $\tau_\ell ({\bm k}) \propto k$, is a very popular model in the graphene literature. Early on it was understood~\cite{chargedimpurities} that such functional dependence of $\tau_\ell$ on $k$ is needed to explain the linear-in-carrier-density d.c. conductivities that are experimentally measured in samples on dielectric substrates such as ${\rm SiO}_2$ (and is typically attributed to charged impurities located close to the graphene sheet).  Below, we will not assume any specific functional dependence of $\tau_\ell$ on $k$. Our low-temperature analytical results for the Coulomb drag transresistivity do {\it not} depend on the particular scattering model that one chooses.

At this point, it is useful to introduce the one-particle spectral function corresponding to Eq.~(\ref{eq:finitelifetimeeffect}),
\ber
\label{eq:spectral_lambda}
{\cal A}_{\ell,\lambda}({\bm k} , \omega) &=& - \frac{1}{\pi} \Im m~{\cal G}^{\rm R}_{\ell, \lambda}({\bm k}, \omega) \nonumber \\
&=& i \left[ {\cal G}^{\rm A}_{\ell, \lambda} ({\bm k} , \omega ) - {\cal G}^{\rm R}_{\ell, \lambda}({\bm k } , \omega)\right]~,
\eer
and the identity
\ber
{\cal G}^{\rm R}_{\ell, \lambda}({\bm k}, \omega){\cal G}^{\rm A}_{\ell, \lambda}({\bm k}, \omega) 
= \tau_\ell({\bm k}) {\cal A}_{\ell,\lambda}({\bm k} , \omega)~.
\eer
Using these definitions in Eq.~(\ref{eq:gamma_diagonal}) we obtain the following approximate expression for the non-linear susceptibility:
\begin{widetext}
\ber\label{eq:gamma_pol1}
\Gamma_\ell({\bm q} , \omega^+, \omega^-) &\approx& vN_{{\rm f}, \ell}\int_{-\infty}^{+\infty}\frac{d \varepsilon}{2\pi i } \int \frac{d^2 {\bm k}}{(2\pi)^2} \sum_{\lambda, \lambda'} \Big\{ \big[ n_{{\rm F}}(\varepsilon + \omega ) - n_{\rm F}(\varepsilon)\big] i \tau_\ell({\bm k} + {\bm q}) 
{\cal A}_{\ell, \lambda}({\bm k} , \varepsilon ) {\cal A}_{\ell, \lambda'} ({\bm k} + {\bm q} , \varepsilon + \omega )
\nonumber\\
&\times& \frac{1 + \lambda \lambda' \cos (\varphi_{{\bm k}} - \varphi_{{\bm k} + {\bm q}})}{2}  \sigma^x _{{\bm k} + {\bm q}\lambda', {\bm k} + {\bm q}, \lambda'}\Big\} 
+ \Big\{ ({\bm q}, \omega) \rightarrow (-{\bm q} , -\omega) \Big\}~.
\eer
\end{widetext}
From now on we will introduce the simplified notation $ J^x _{{\bm k}, \lambda} \equiv v \sigma^x_{{\bm k} \lambda, {\bm k} \lambda} = v \lambda \cos (\varphi_{\bm k})$, where in the last equality we have used Eq.~(\ref{eq:sigma_vertex}). The definition of $J^x _{{\bm k}, \lambda}$ should not be confused with the definition of the current-density operator ${\hat J}^{i}_{{\bm q}, \ell}$ in second quantization given in Eq.~(\ref{eq:currentoperator}).

In this Article we are interested in the limit in which intralayer scattering is weak, which is, for example, the most relevant regime for high-quality DLG samples. In this limit we can approximate the spectral functions in Eq.~(\ref{eq:gamma_pol1}) with $\delta$ functions,
\be\label{eq:spectralfunctiondeltalike}
{\cal A}_{\ell, \lambda}({\bm k} , \varepsilon ) \approx \delta(\omega - \xi^{(\ell)}_{{\bm k}, \lambda})~.
\ee

Straightforward algebraic manipulations of Eq.~(\ref{eq:gamma_pol1}) with the use of Eq.~(\ref{eq:spectralfunctiondeltalike}) yield the following Boltzmann-transport (BT) expression for the non-linear susceptibility:
\begin{widetext}
\ber\label{eq:gamma_pol2}
\Gamma^{\rm BT}_\ell ({\bm q}, \omega^+, \omega^-) &\equiv& \Gamma^{\rm BT}_\ell ({\bm q}, \omega)  = N_{{\rm f}, \ell}\sum_{\lambda , \lambda'} \int \frac{d^2 {\bm k}}{(2\pi)^2} \big[\tau_\ell ({\bm k} + {\bm q} ) J^x_{{\bm k} + {\bm q} , \lambda}  - \tau_\ell ({\bm k}) J^x _{{\bm k} , \lambda} \big]\nonumber \\
&\times&\Im m \Bigg\{ \frac{n_{{\rm F}} (\xi^{(\ell)}_{{\bm k}, \lambda}) - n_{{\rm F}} (\xi^{(\ell)} _{{\bm k} + {\bm q}, \lambda'})}{\omega + \xi^{(\ell)}_{{\bm k} , \lambda} -\xi^{(\ell)} _{{\bm k} + {\bm q} , \lambda'} + i 0^+}
\big[1 + \lambda \lambda ' \cos (\varphi_{{\bm k} +{\bm q}} - \varphi_{{\bm k}})\big]\Bigg\}~.\nonumber\\
\eer
\end{widetext}
We have dubbed Eq.~(\ref{eq:gamma_pol2}) ``BT expression" for the non-linear susceptibility since after inserting it in 
Eq.~(\ref{eq:conductivity_end}) one obtains precisely an expression for the Coulomb drag conductivity that can be derived 
within a Boltzmann-equation approach. 
This will be shown in Appendix~\ref{appendix:boltzmann}.

Note that, in the usual ``relaxation-time approximation" ($\tau_\ell$ independent of $k$), the BT expression for the non-linear susceptibility is simply proportional to the intralayer scattering time, $\Gamma^{\rm BT}_\ell ({\bm q}, \omega)  \propto \tau_\ell$.

\subsection{Coulomb drag transresistivity in the Boltzmann-transport limit}
\label{sect:rho_drag}

We now focus our attention on the quantity which is actually measured in experiments, {\it i.e.} the drag transresistivity $\rho_{\rm D}$, which is precisely the ratio between the voltage drop in the passive layer and the current in the active layer is indeed $\rho_{\rm D}$. This quantity can be easily found by inverting the $2\times 2$ conductivity matrix,
\be
\label{eq:remind_rhodrag}
\rho_{{\rm D}} = - \sigma_{\rm D} \frac{1}{\displaystyle 
{\rm det}
\left(
\begin{array}{cc}
\sigma_1 & \sigma_{\rm D}\\
\sigma_{\rm D} & \sigma_2
\end{array}
\right)} \approx -\frac{\sigma_{\rm D}}{\sigma_1\sigma_2}~,
\ee
where $\sigma_\ell$ is the intralayer conductivity and the last approximation in Eq.~(\ref{eq:remind_rhodrag}) holds true only if $\sigma_{\rm D} \ll \sigma_\ell$.
Within BT theory the intralayer conductivity at finite temperature is given by~\cite{dassarma_rmp_2011}
\be\label{eq:cond_intra_finiteT}
\sigma_\ell = \frac{e^2}{2}v^2 N_{{\rm f}, \ell} \sum_{\lambda}\int \frac{d^2{\bm k}}{(2\pi)^2}\tau_\ell({\bm k})\Bigg[-\frac{\partial n_{\rm F}(\xi^{(\ell)}_{{\bm k}, \lambda})}{\partial \xi^{(\ell)}_{{\bm k}, \lambda}}\Bigg]~,
\ee
while the drag conductivity $\sigma_{\rm D}$ is given by Eq.~(\ref{eq:conductivity_end}) with the expression (\ref{eq:gamma_pol2}) for the non-linear susceptibility. 
The final expression for the BT drag resistivity reads
\begin{widetext}
\ber\label{eq:rho_general_finiteT}
\rho^{\rm BT}_{\rm D} &=& - \frac{\beta e^2}{16\pi \sigma_1\sigma_2} \int \frac{d^2 {\bm q}}{(2\pi)^2} \int_{-\infty}^{+\infty} d \omega \frac{|U_{12} (q,\omega)|^2}{\sinh^2(\beta \omega/2)} \Gamma^{\rm BT}_1({\bm q},\omega) \Gamma^{\rm BT}_2({\bm q}, \omega) ~. 
\eer
\end{widetext}
Note that in the relaxation-time approximation the BT drag transresistivity does not depend on the transport scattering times in the two layers (even if these are different), since, as noted earlier in Sect.~\ref{sect:boltzmann}, $\Gamma^{\rm BT}_\ell$ is proportional to $\tau_\ell$ in this approximation and so is $\sigma_\ell$. 

The situation turns out to be much more complicated in the case in which $\tau_\ell$ is an arbitrary function of $k = |{\bm k}|$. A full numerical treatment of the three-dimensional integral in Eq.~(\ref{eq:rho_general_finiteT}) at any finite temperature is beyond the scope of the present work and will be the subject of a forthcoming publication.

In the next Section, however, we demonstrate that a dramatic simplification occurs in the low-temperature limit. We will indeed prove that when $k_{\rm B} T$ is much smaller than the Fermi energies in the two layers, $k_{\rm B} T \ll \min_\ell(\varepsilon_{{\rm F}, \ell})$, $\rho^{\rm BT}_{\rm D}$ is {\it insensitive} to the precise functional dependence $\tau_\ell  = \tau_{\ell}(k)$ of the intralayer scattering time. Since for typical values of doping the Fermi energy in a graphene sheet is very large ($\approx 1400~{\rm K}$ for a carrier density on the order of $10^{12}~{\rm cm}^{-2}$), analytical results in the low-temperature limit are very useful when both layers are sufficiently away from the charge neutrality point. 

In the intriguing regime in which one layer (both layers) lies (lie) at the charge neutrality point BT theory is inapplicable. 
Coulomb drag in this regime is dominated by unavoidable electron-hole puddles~\cite{STSSTM} and is certainly very interesting but well beyond the scope of the present Article. For reasons of symmetry, lowest-order Boltzmann transport theory applied to the regime in which one of the two layers is at the neutrality point gives $\rho^{\rm BT}_{\rm D} =0$~\cite{tse_prb_2007,narozhny_prb_2007}.

\section{The low-temperature limit}
\label{sect:lowT}

In this Section we use Eq.~(\ref{eq:rho_general_finiteT}) to calculate $\rho^{\rm BT}_{\rm D}$ {\it analytically} in the low-temperature limit, 
for arbitrary values of the dielectric constants $\epsilon_i$ in Fig.~\ref{fig:one}, and for a generic scattering time $\tau_\ell(k)$.

As already anticipated above, the low-temperature regime is readily identified by the inequality 
$k_{{\rm B}} T \ll \min_\ell (\varepsilon_{{\rm F},\ell})$.  In this limit we can: 

(i) use the low-temperature expression for the intralayer conductivity~\cite{chargedimpurities,dassarma_rmp_2011}
\be\label{eq:cond_intra}
\lim_{T \to 0} \sigma_\ell = \frac{e^2}{4\pi}~N_{{\rm f}, \ell} \varepsilon_{{\rm F}, \ell} 
\tau_\ell(k_{{\rm F}, \ell})~;
\ee

ii) evaluate the non-linear BT susceptibilities
$\Gamma^{\rm BT}_1({\bm q},\omega)$ and $\Gamma^{\rm BT}_2({\bm q},\omega)$ at zero temperature and only to lowest order in $\omega$ in the low-frequency $\omega/\min_\ell (\varepsilon_{{\rm F},\ell}) \to 0$ limit; 

and iii) replace the dynamically-screened interlayer interaction with the much simpler statically-screened interaction $U_{12}(q, 0)$. 

Indeed, the thermal factor $\sinh^{-2}(\beta\omega/2)$ in Eq.~(\ref{eq:rho_general_finiteT}) represents an effective cut off on the values of $\omega$ in the integral in Eq.~(\ref{eq:rho_general_finiteT}): it must be $\omega \ll 2/\beta$. 

To find the asymptotic behavior of $\Gamma^{\rm BT}_\ell ({\bm q}, \omega)$ in the limit $\omega \to 0$ we first re-write Eq.~(\ref{eq:gamma_pol2}) in the form
\begin{widetext}
\ber\label{eq:gamma_pol2_rewritten}
\Gamma^{\rm BT}_\ell ({\bm q}, \omega)  &=& -\pi N_{{\rm f}, \ell}\sum_{\lambda , \lambda'} \int \frac{d^2 {\bm k}}{(2\pi)^2} 
\big[\tau_\ell ({\bm k} + {\bm q} ) J^x_{{\bm k} + {\bm q} , \lambda}  - \tau_\ell ({\bm k}) J^x _{{\bm k} , \lambda} \big]
\big[n_{\rm F}(\xi^{(\ell)}_{{\bm k}, \lambda}) - n_{\rm F}(\xi^{(\ell)}_{{\bm k} + {\bm q}, \lambda'})\big]
\nonumber \\
&\times& 
\delta(\omega + \xi^{(\ell)}_{{\bm k} , \lambda} -\xi^{(\ell)} _{{\bm k} + {\bm q} , \lambda'})
\big[1 + \lambda \lambda ' \cos (\varphi_{{\bm k} +{\bm q}} - \varphi_{{\bm k}})\big]~.\nonumber\\
\eer
\end{widetext}
Clearly $\Gamma^{\rm BT}_\ell ({\bm q}, 0) =0$. 
We thus need to calculate the derivative of $\Gamma^{\rm BT}_\ell ({\bm q}, \omega)$ with respect to $\omega$ at $\omega=0$ and in the zero-temperature limit. We find
\begin{widetext}
\ber
\left.\frac{\partial \Gamma^{\rm BT}_\ell ({\bm q}, \omega)}{\partial \omega}\right|_{\omega =0}  &=& - \pi N_{{\rm f}, \ell}\sum_{\lambda , \lambda'} \int \frac{d^2 {\bm k}}{(2\pi)^2} 
\big[\tau_\ell ({\bm k} + {\bm q} ) J^x_{{\bm k} + {\bm q} , \lambda}  - \tau_\ell ({\bm k}) J^x _{{\bm k} , \lambda} \big]
~\left.\frac{\partial \big[n_{\rm F}(\omega + \xi^{(\ell)}_{{\bm k} , \lambda})\big]}{\partial\omega}
\right|_{\omega =0}
\nonumber \\
&\times& 
\delta(\xi^{(\ell)}_{{\bm k} , \lambda} -\xi^{(\ell)} _{{\bm k} + {\bm q} , \lambda'})
\big[1 + \lambda \lambda ' \cos (\varphi_{{\bm k} +{\bm q}} - \varphi_{{\bm k}})\big] \nonumber\\
&\stackrel{T \to 0}{=} &
- \pi N_{{\rm f}, \ell} \sum_{\lambda , \lambda'} \int \frac{d^2 {\bm k}}{(2\pi)^2} 
\big[\tau_\ell ({\bm k} + {\bm q} ) J^x_{{\bm k} + {\bm q} , \lambda}  - \tau_\ell ({\bm k}) J^x _{{\bm k} , \lambda} \big]
\delta(\xi^{(\ell)}_{{\bm k} , \lambda})
\delta(\xi^{(\ell)}_{{\bm k} , \lambda} -\xi^{(\ell)} _{{\bm k} + {\bm q} , \lambda'}) \nonumber\\
&\times&
\big[1 + \lambda \lambda ' \cos (\varphi_{{\bm k} +{\bm q}} - \varphi_{{\bm k}})\big]~,\nonumber\\
\eer
\end{widetext}
where the last equality is valid only in the zero-temperature limit. More explicitly, we find
\begin{widetext}
\ber\label{eq:gammalowT}
\lim_{T \to 0} \Gamma^{\rm BT}_\ell ({\bm q}, \omega\to 0) &=& 
- \pi \omega N_{{\rm f}, \ell} \sum_{\lambda , \lambda'} \int \frac{d^2 {\bm k}}{(2\pi)^2} 
\big[\tau_\ell ({\bm k} + {\bm q} ) J^x_{{\bm k} + {\bm q} , \lambda}  - \tau_\ell ({\bm k}) J^x _{{\bm k} , \lambda} \big]
\delta(\xi^{(\ell)}_{{\bm k} , \lambda})
\delta(\xi^{(\ell)} _{{\bm k} + {\bm q} , \lambda'}) \nonumber\\
&\times&
\big[1 + \lambda \lambda ' \cos (\varphi_{{\bm k} +{\bm q}} - \varphi_{{\bm k}})\big]~.
\eer
\end{widetext}
Eq.~(\ref{eq:gammalowT}) is one of the most important results of this Section. An explicit expression for  $\lim_{T \to 0} \Gamma^{\rm BT}_\ell ({\bm q}, \omega\to 0)$ will be given below in Eq.~(\ref{eq:finalresultGammalowT}). Notice that the two $\delta$-functions in the above expression impose that only intraband excitations ($\lambda = \lambda' = +1$) contribute to $\Gamma^{\rm BT}_\ell ({\bm q}, \omega\to 0)$ in the low-temperature limit. Moreover, the two $\delta$-functions together pin the absolute values of ${\bm k}$ and ${\bm k} + {\bm q}$ to be equal to $k_{{\rm F}, \ell}$. This implies that in the low-temperature limit the transport scattering times $\tau_\ell ({\bm k} + {\bm q})$ and $\tau_\ell ({\bm k})$ inside the square bracket in Eq.~(\ref{eq:gammalowT}) must be both evaluated on the Fermi surface of the $\ell$-th layer, {\it i.e.} $|{\bm k} +{\bm q}| = |{\bm k}| = k_{{\rm F}, \ell}$. Since $\tau_\ell({\bm k})$ depends {\it only} on the absolute value of its argument (and not on the polar angle of ${\bm k}$), {\it both} $\tau_\ell ({\bm k} + {\bm q})$ {\it and} $\tau_\ell ({\bm k})$ {\it can be factorized out of the integral in Eq.~(\ref{eq:gammalowT})}. This is precisely the reason why at the end of Sect.~\ref{sect:rho_drag} we claimed that in the low-temperature limit $\rho^{\rm BT}_{\rm D}$ is completely insensitive to the precise intralayer scattering mechanism.

More mathematically, we have:
\ber\label{eq:gammalowTsimpler}
\lim_{T \to 0} \Gamma^{\rm BT}_\ell ({\bm q}, \omega\to 0) &=& 
- \pi \omega N_{{\rm f}, \ell} \tau_{\ell}(k_{{\rm F}, \ell}) \nonumber\\
&\times&\int_0^{+\infty} \frac{k dk}{2\pi}\delta(v k - \varepsilon_{{\rm F}, \ell})\nonumber\\
&\times&
\int_0^{2\pi}\frac{d \varphi_{\bm k}}{2\pi} \big[J^x_{{\bm k} + {\bm q}, +}  - J^x _{{\bm k},+} \big]\nonumber\\
&\times&
\delta(vk  - v |{\bm k}+{\bm q}|) \nonumber\\
&\times&
\big[1 + \cos (\varphi_{{\bm k} +{\bm q}} - \varphi_{{\bm k}})\big]~.
\eer
We now employ the following identity,
\be
J^x _{{\bm k} + {\bm q}, +} - J^x _{{\bm k},+} = \frac{q}{k_{{\rm F}, \ell}}~\cos (\varphi_{ {\bm q}})~,
\ee
which can be easily derived by using the aforementioned condition $|{\bm k} +{\bm q}| = |{\bm k}| = k_{{\rm F},\ell}$. 
The two integrals in Eq.~(\ref{eq:gammalowTsimpler}) can be easily carried out analytically: we find
\ber\label{eq:finalresultGammalowT}
\lim_{T \to 0} \Gamma^{\rm BT}_\ell ({\bm q}, \omega\to 0) &=& - N_{{\rm F}, \ell}\frac{\omega q_x \tau_\ell(k_{{\rm F},\ell})}{2\pi v q}\Theta(2k_{{\rm F}, \ell} -q)  \nonumber \\
&\times&\sqrt{1- \frac{q^2}{4 k^2_{{\rm F}, \ell}}}~,
\eer
where we have used that $q_x = q \cos(\varphi_{{\bm q}})$. Eq.~(\ref{eq:finalresultGammalowT}) is one of the most important results of this Section. We stress again that the applicability of this equation is not limited to the case of a momentum-independent scattering time. Rather,  Eq.~(\ref{eq:finalresultGammalowT}) applies to a generic intralayer scattering time $\tau_\ell = \tau_\ell(k)$. The physical meaning of this result is that in the low-temperature limit the non-linear susceptibility is determined by what happens close to the Fermi surface. What matters thus is only the magnitude of $\tau_\ell(k)$ evaluated at the Fermi momentum $k_{{\rm F}, \ell}$ of the $\ell$-th layer. This conclusion is in agreement with Narozhny {\it et al.}~\cite{narozhny_arXiv_2011}.

Using Eq.~(\ref{eq:finalresultGammalowT}) and the following integral
\be
\int_{-\infty}^{+\infty} dx~\frac{x^2}{\sinh^2 (y x/2)} = \frac{8 \pi^2}{3 y^3}~,
\ee
we finally arrive at the desired result for the low-temperature BT drag transresistivity:
\begin{widetext}
\ber\label{eq:CDgenerallowT}
\lim_{T \to 0}\rho^{\rm BT}_{\rm D} = - \frac{(k_{\rm B}T)^2}{ 6 e^2 \varepsilon_{{\rm F},1} \varepsilon_{{\rm F},2} v^2} \int_0^{q_{\rm max}} dq~ q|U_{12} (q, 0)|^2
\sqrt{1- \frac{q^2}{4 k^2_{{\rm F}, 1}}}\sqrt{1- \frac{q^2}{4 k^2_{{\rm F}, 2}}} ~,
\eer
\end{widetext}
where $q_{\rm max} \equiv \min(2k_{{\rm F},1}, 2k_{{\rm F},2})$. Once again, we stress that the result in Eq.~(\ref{eq:CDgenerallowT}) does {\it not} depend on the precise functional form of $\tau_\ell(k)$.

\subsection{Dimensionless variables}
\label{sect:dimensionless}

It is now useful to introduce dimensionless variables. We scale the wave number $q$ in Eq.~(\ref{eq:CDgenerallowT}) with $\sqrt{k_{{\rm F}, 1} k_{{\rm F}, 2}}$ by introducing $x = q/\sqrt{k_{{\rm F}, 1} k_{{\rm F}, 2}}$ and the effective interaction $U_{12}$ with $e^2/\sqrt{k_{{\rm F}, 1} k_{{\rm F}, 2}}$ by introducing ${\bar U}_{12} = U_{12}\sqrt{k_{{\rm F}, 1} k_{{\rm F}, 2}}/e^2$. 

In these reduced units Eq.~(\ref{eq:CDgenerallowT}) reads (for physical reasons we restore Planck's constant from now on)
\begin{widetext}
\ber\label{eq:CDgenerallowTdimensionless}
\lim_{T \to 0}\rho^{\rm BT}_{\rm D} = - \frac{h}{e^2}~\frac{\alpha^2_{\rm ee}}{ 12\pi }\frac{(k_{\rm B}T)^2}{\varepsilon_{{\rm F},1} \varepsilon_{{\rm F},2}} \int_0^{x_{\rm max}} dx~ x|{\bar U}_{12} (x, 0)|^2
\sqrt{1- \frac{x^2}{4}\frac{k_{{\rm F}, 1}}{k_{{\rm F}, 2}}}
\sqrt{1- \frac{x^2}{4}\frac{k_{{\rm F}, 2}}{k_{{\rm F}, 1}}}
~,
\eer
\end{widetext}
where $x_{\rm max} = \min(2\sqrt{k_{{\rm F}, 1}/k_{{\rm F}, 2}}, 2\sqrt{k_{{\rm F}, 2}/k_{{\rm F}, 1}})$. In Eq.~(\ref{eq:CDgenerallowTdimensionless}) we have introduced the dimensionless coupling constant $\alpha_{\rm ee} = e^2/(\hbar v)$, which has a value $\approx 2.2$ in DLG and $  \approx 4.4$ in Bi$_2$Te$_3$ TIs if we use the respective Dirac velocities $v_{\rm G} \approx 10^{6}~{\rm m}/{\rm s}$ and $v_{\rm TI} \approx 5 \times 10^{5}~{\rm m}/{\rm s}$. Eq.~(\ref{eq:CDgenerallowTdimensionless}) is the most important result of this Section.

To proceed further, we write in an explicit manner the statically-screened dimensionless interlayer interaction ${\bar U}_{12} (x, 0)$:
\begin{widetext}
\ber\label{eq:dimensionlessinteraction}
{\bar U}_{12}(x, 0) = \frac{ 8 \pi x f_{12}(x\xi )}{[x + 2 N_{{\rm f}, 1}\alpha_{\rm ee} \sqrt{k_{{\rm F}, 1}/k_{{\rm F}, 2}}~f_{11}(x \xi)]
[x + 2 N_{{\rm f}, 2}\alpha_{\rm ee}\sqrt{k_{{\rm F}, 2}/k_{{\rm F}, 1}}~f_{22}(x\xi)] - 16  N_{{\rm f}, 1}N_{{\rm f}, 2}\alpha^2_{\rm ee}f_{12}^2 (x \xi)}~,
\eer
\end{widetext}
where we have introduced the crucially-important dimensionless parameter
\be
\xi \equiv d \sqrt{k_{{\rm F},1} k_{{\rm F},2}}~,
\ee
and the dimensionless form factors
\be
\label{eq:f11}
f_{11}(x) = \frac{\left(\epsilon_2 + \epsilon_3 \right) e^{x} + \left(\epsilon_2 - \epsilon_3 \right) e^{-x}}{g(x)} ~,
\ee
\be
\label{eq:f22}
f_{22}(x) = \frac{\left(\epsilon_2 + \epsilon_1\right)e^{x} + \left(\epsilon_2 - \epsilon_1 \right) e^{-x}}{g(x)}~,
\ee
\be
\label{eq:f12}
f_{12}(x) = \frac{\epsilon_2}{g(x)}~,
\ee
with 
\be
\label{eq:gx}
g(x) = (\epsilon_1 + \epsilon_2)(\epsilon_2 + \epsilon_3) e^{x} + 
(\epsilon_1 - \epsilon_2)(\epsilon_2 - \epsilon_3)e^{-x}~.
\ee
The only ingredient we have used to write Eq.~(\ref{eq:dimensionlessinteraction}) is the well-known behavior of the static Lindhard function of a doped MD2DES for wave numbers $q$ smaller than twice the Fermi momentum $k_{{\rm F}, \ell}$:
\be
\chi^{(0)}_\ell(q \leq 2 k_{{\rm F}, \ell}, 0) = - \frac{N_{{\rm f},\ell} k_{{\rm F}, \ell}}{2\pi \hbar v}~.
\ee

Two asymptotic limits can now be easily inferred from the general low-temperature theory in Eqs.~(\ref{eq:CDgenerallowTdimensionless})-(\ref{eq:gx}): 1) the weak-coupling (``large interlayer distance" and/or ``high density") limit, {\it i.e.} the limit in which $\xi \gg 1$, and 2) the strong-coupling (``small interlayer distance" and/or  ``low density") limit, {\it i.e.} the limit in which $\xi \ll 1$. 

As discussed briefly above, the applicability of BT theory in the low-density regime is questionable: for the BT theory to remain valid, one should think of entering the strong-coupling regime by reducing the interlayer separation $d$ while keeping fixed the factor $\sqrt{k_{{\rm F},1} k_{{\rm F},2}}$. Note also that, in the strong-coupling regime, diagrams of higher order with respect to those presented in Fig.~\ref{fig:two} might become relevant. The applicability of RPA -- Eq.~(\ref{eq:dynamicalRPA}) -- is also questionable in this regime. Only a comparison with experimental results can tell us about the importance of these subtle issues of many-body theory.

\subsection{The weak-coupling limit}
\label{sect:weak}

In the limit $\xi \gg 1$, straightforward algebraic manipulations of Eqs.~(\ref{eq:CDgenerallowTdimensionless})-(\ref{eq:gx}) yield the following result for the low-temperature BT drag transresistivity:
\ber \label{eq:rhoD_GG_weakcoupling}
\lim_{\xi \to \infty}\lim_{T \to 0}\rho^{\rm BT}_{\rm D} &=& - \frac{h}{e^2}~\frac{\pi}{3}~\frac{(k_{\rm B} T)^2}{\varepsilon_{{\rm F},1} \varepsilon_{{\rm F},2} \xi^4 N^2_{{\rm f}, 1}N^2_{{\rm f}, 2}\alpha^2_{\rm ee}} \nonumber \\
&\times&\int_0 ^{+\infty} dy ~ \frac{y^3 f^2_{12}(y)}{[f_{11}(y) f_{22}(y) - 4 f^2_{12}(y)]^2}~,\nonumber \\
\eer
where we have introduced a new reduced variable, $y = x\xi$. The quadrature with respect to the variable $y$ can be carried out analytically yielding the following result:
\ber
&&\int_0 ^{+\infty} dy~\frac{y^3 f^2_{12}(y)}{[f_{11}(y) f_{22}(y) - 4 f^2_{12}(y)]^2} \nonumber\\
&=&\frac{1}{4}\epsilon^2_2\int_0 ^{+\infty} dy~\frac{y^3}{\sinh^2(y)} = \frac{3}{8}\zeta(3)\epsilon^2_2~,
\eer
where $\zeta(x)$ is the Riemann zeta function and $\zeta(3) \approx 1.2$. Note that this quadrature does not depend neither on $\epsilon_1$ nor on $\epsilon_3$.

In summary, we find
\ber\label{eq:rhoD_GG_weakcoupling_final}
\lim_{\xi \to \infty}\lim_{T \to 0}\rho^{\rm BT}_{\rm D} &=& -\frac{h}{e^2}~\frac{\pi  \zeta (3)}{8}~\frac{\epsilon_2 ^2}{d^4 \alpha^2_{\rm ee}  N^2_{{\rm f},1} N^2_{{\rm f},2} k^2_{{\rm F}, 1}k^2_{{\rm F}, 2}}\nonumber\\
&\times&\frac{(k_{\rm B}T)^2}{\varepsilon_{{\rm F},1}\varepsilon_{{\rm F},2}} \propto - \frac{h}{e^2}~\epsilon_2 ^2~\frac{(k_{\rm B} T)^2}{n_1 ^{3/2} n_2^{3/2} d^4}~.\nonumber\\
\eer
The dependence of $\rho^{\rm BT}_{\rm D}$ on temperature, densities and interlayer distance displayed by Eq.~(\ref{eq:rhoD_GG_weakcoupling_final}) is in agreement with the results by Tse {\it et al.}~\cite{tse_prb_2007} and by Katsnelson~\cite{katsnelson_prb_2011}, even though the numerical prefactor we find is four times smaller (we remind the reader that for DLG $N_{{\rm f}, 1} = N_{{\rm f}, 2} =4$). It is important to observe that the weak-coupling and low-temperature BT Coulomb drag transresistivity is sensitive only to the dielectric constant $\epsilon_2$ between the two MD2DESs (see Fig.~\ref{fig:one}).

\subsection{The strong-coupling limit}
\label{sect:strong}

In the limit $\xi \to 0$ we can expand all the functions $f_{ij}(x\xi)$ that appear in the effective interaction (\ref{eq:dimensionlessinteraction}) is powers of their argument since $x$ is bounded from above, $0 \leq x \leq x_{\rm max}$, and $\xi \to 0$. 
Following this procedure we find the following asymptotic expression for ${\bar U}_{12}(x,0)$ in the limit $\xi \to 0$:
\be
\lim_{\xi \to 0} {\bar U}_{12}(x,0) = \frac{4\pi}{\displaystyle (\epsilon_1 + \epsilon_3 )x + 2 \frac{q_{{\rm TF},1} + q_{{\rm TF},2}}{\sqrt{k_{{\rm F},1} k_{{\rm F},2}}}}~,
\ee
where we have introduced the Thomas-Fermi screening wave number $q_{{\rm TF}, \ell} = N_{{\rm f}, \ell} \alpha_{\rm ee} k_{{\rm F}, \ell}$.
Using this result in Eq.~(\ref{eq:CDgenerallowTdimensionless}) we find the final expression for the low-temperature BT drag transresistivity in the strong-coupling  limit:
\ber\label{eq:rhoD_GG_strongcoupling}
\lim_{\xi \to 0} \lim_{T \to 0}\rho^{\rm BT}_{\rm D} &=& - \frac{h}{e^2}~\frac{4 \pi }{3}~\alpha_{{\rm ee}}^2 \frac{(k_{\rm B} T)^2}{\varepsilon_{{\rm F},1}\varepsilon_{{\rm F},2}} \nonumber \\
&\times&\int_0 ^{x_{{\rm max}}}dx~\sqrt{1- \frac{x^2}{4}\frac{k_{{\rm F}, 1}}{k_{{\rm F}, 2}}}
\sqrt{1- \frac{x^2}{4}\frac{k_{{\rm F}, 2}}{k_{{\rm F}, 1}}}\nonumber\\
&\times& \frac{x}{\displaystyle \Bigg[(\epsilon_1 + \epsilon_3 )x + 2 \frac{q_{{\rm TF},1} + q_{{\rm TF},2}}
{\sqrt{k_{{\rm F},1} k_{{\rm F},2}}}\Bigg]^2}~.\nonumber\\
\eer
Eq.~(\ref{eq:rhoD_GG_strongcoupling}) simplifies considerably if we assume equal densities and degeneracies in the two layers:
$n_1 =n_2 \equiv n$, $k_{{\rm F},1} = k_{{\rm F},2} \equiv k_{\rm F}$, $\varepsilon_{{\rm F},1} = \varepsilon_{{\rm F},2} \equiv \varepsilon_{\rm F}$, and $N_{{\rm f}, 1} = N_{{\rm f}, 2} \equiv N_{\rm f}$. In this case Eq.~(\ref{eq:rhoD_GG_strongcoupling}) reduces to
\be\label{eq:rhoD_GG_strongcoupling_equallayers}
\lim_{\xi \to 0} \lim_{T \to 0}\rho^{\rm BT}_{\rm D} = - \frac{h}{e^2}~\frac{4 \pi }{3}~\alpha_{{\rm ee}}^2 \left(\frac{k_{\rm B} T}{\varepsilon_{\rm F}}\right)^2{\cal F}(\epsilon_1+\epsilon_3,N_{\rm f}\alpha_{\rm ee})~,
\ee
where
\be\label{eq:strongcouplingintegral}
{\cal F}(\epsilon_1+\epsilon_3,N_{\rm f}\alpha_{\rm ee}) \equiv \int_0 ^2dx~\frac{x (1-x^2/4)}{\big[(\epsilon_1 + \epsilon_3 )x + 4 N_{\rm f}\alpha_{\rm ee}\big]^2}~.
\ee
Note that in the strong-coupling limit $\lim_{T \to 0}\rho^{\rm BT}_{\rm D}$ does not depend on the interlayer distance $d$ 
and scales like $1/n$. As far as the dielectric constants $\epsilon_i$ are concerned, the strong-coupling and low-temperature BT Coulomb drag transresistivity depends only on $\epsilon_1+\epsilon_3$ but not on $\epsilon_2$.

The quadrature in Eq.~(\ref{eq:strongcouplingintegral}) can be easily carried out analytically. The result is
\ber\label{eq:Ffunction}
{\cal F}(\epsilon_1+\epsilon_3,N_{\rm f}\alpha_{\rm ee}) &=&
\frac{1}{2(\epsilon_1 + \epsilon_3)^4} \Bigg\{ 12 N_{\rm f}\alpha_{\rm ee}(\epsilon_1+\epsilon_3) \nonumber\\
&-& (\epsilon_1+\epsilon_3)^2[3 + 2 \ln(2)] + 24 N^2_{\rm f} \alpha^2_{\rm ee} \nonumber\\
&\times&\ln(2) + 2[(\epsilon_1+\epsilon_3)^2 -12 N^2_{\rm f}\alpha^2_{\rm ee}]\nonumber\\
 &\times&\ln{\left(2+ \frac{\epsilon_1+\epsilon_3}{N_{\rm f}\alpha_{\rm ee}}\right)} \Bigg\}~.
\eer
Note that ${\cal F}(\epsilon_1+\epsilon_3,N_{\rm f}\alpha_{\rm ee})$ diverges logarithmically in the weak-screening $N_{\rm f} \alpha_{\rm ee} \to 0$ limit, in agreement with Eq.~(41) of Ref.~\onlinecite{narozhny_arXiv_2011}.

\section{Deviations from the low-temperature quadratic behavior due to a frequency-dependent substrate dielectric constant}
\label{sect:epsilon3omega}

In this Section we illustrate how deviations from the Fermi-liquid quadratic-in-temperature dependence of the Coulomb drag transresistivity can occur in the situation in which the dielectric constant $\epsilon_i$ of one of the media surrounding the two Dirac-fermion layers has a strong frequency dependence. 

Consider, for example, a DLG system on a substrate like ${\rm SrTiO}_3$. Following Ref.~\onlinecite{han_apl_2007}, we model the dielectric constant of this substrate by a frequency-dependent function of the form
\be\label{eq:epsilon3omega}
\epsilon_3 (\omega) = \epsilon_{\infty} + \left(\epsilon_{0} - \epsilon_{\infty}\right) \frac{\omega_0 ^2}{\omega_0 ^2 - \omega^2 + i \gamma \omega}~.
\ee
At room temperature $\epsilon_{\infty} = 5.2$ , $\epsilon_0 = 310$, $\omega_0 /(2\pi) = 2.7~{\rm THz}$, and $\gamma/(2\pi) = 1.3~{\rm THz}$.

We would like to understand what is the qualitative role played by the change $\epsilon_3 \to \epsilon_3(\omega)$ in the Coulomb drag transresistivity. For simplicity, we assume that the two graphene sheets comprising the double layer have the same density $n_1=n_2 =n$ ($k_{{\rm F},1} = k_{{\rm F},2} \equiv k_{{\rm F}}$). We also assume that the Fermi energy $\varepsilon_{\rm F}$ corresponding to $n$ is the largest energy scale in the problem: under this assumption, we are still allowed to expand the BT non-linear susceptibility $\Gamma^{\rm BT}_\ell({\bm q}, \omega)$ to lowest order in the small parameter $\hbar\omega/\varepsilon_{\rm F}$. 

Despite $\hbar\omega \ll \varepsilon_{\rm F}$, we can have two distinct regimes since $\epsilon_3(\omega)$ brings in a new frequency scale, {\it i.e.} $\omega_0$: i) $\omega \ll \omega_0 \ll \varepsilon_{\rm F}/\hbar$ and ii) $\omega_0 \ll \omega \ll \varepsilon_{\rm F}/\hbar$. In the first regime what matters is obviously $\epsilon_3(\omega=0) = \epsilon_0$. In the second regime, instead, we have to retain the full frequency dependence of $\epsilon_3(\omega)$. The existence of this second regime ensures the possibility to observe deviations from $\rho^{\rm BT}_{\rm D} \propto T^2$ above a certain temperature scale.

The low-temperature Coulomb drag transresistivity for the situation described above is given by Eq.~(\ref{eq:rho_general_finiteT}) with $\Gamma^{\rm BT}_\ell({\bm q}, \omega)$ 
given by the expression in Eq.~(\ref{eq:finalresultGammalowT}) and $\sigma_\ell$ given by the expression in Eq.~(\ref{eq:cond_intra}):
\ber\label{eq:dragepsilon3omega}
\lim_{T \to 0}\rho^{\rm BT}_{\rm D} &=& -\frac{h}{e^2} \frac{\hbar \beta}{16 \pi^3 v^2 \varepsilon^2_{\rm F}}\int_0^{2k_{\rm F}} dq~q\Big(1-\frac{q^2}{4 k^2_{\rm F}}\Big) \nonumber\\
&\times& \int_0^{+\infty}d\omega~\frac{\omega^2 \big|\left.U_{12}(q,0)\right|_{\epsilon_3 \to \epsilon_3(\omega)}\big|^2}{\sinh^2(\beta\omega/2)}~.
\eer
Here the notation $\left.U_{12}(q,0)\right|_{\epsilon_3 \to \epsilon_3(\omega)}$ means that from the point of view of the electronic system we still have to use the statically-screened interlayer interaction, while from the point of view of the substrate we have to take into account the frequency dependence of $\epsilon_3$ through the use of Eq.~(\ref{eq:epsilon3omega}). 

\begin{figure}
\centering
\includegraphics[width=1.0\linewidth]{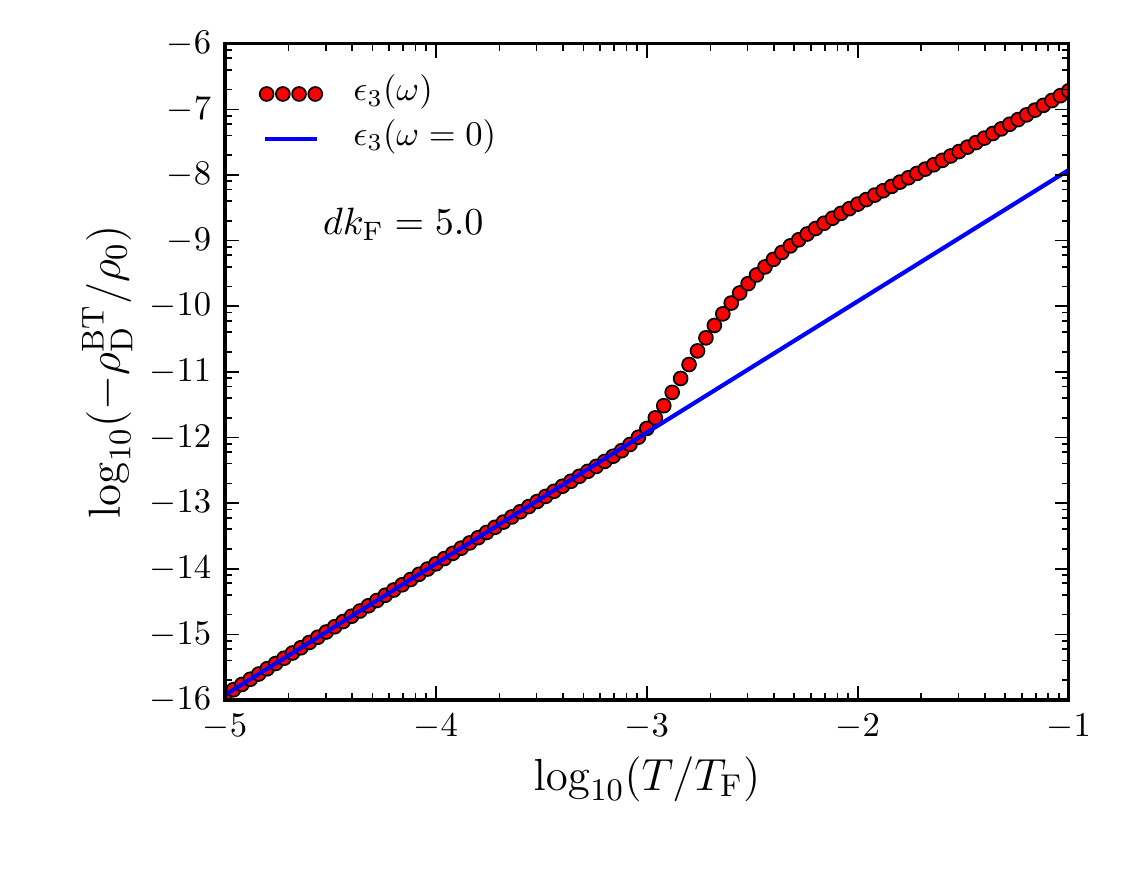}\\
\includegraphics[width=1.0\linewidth]{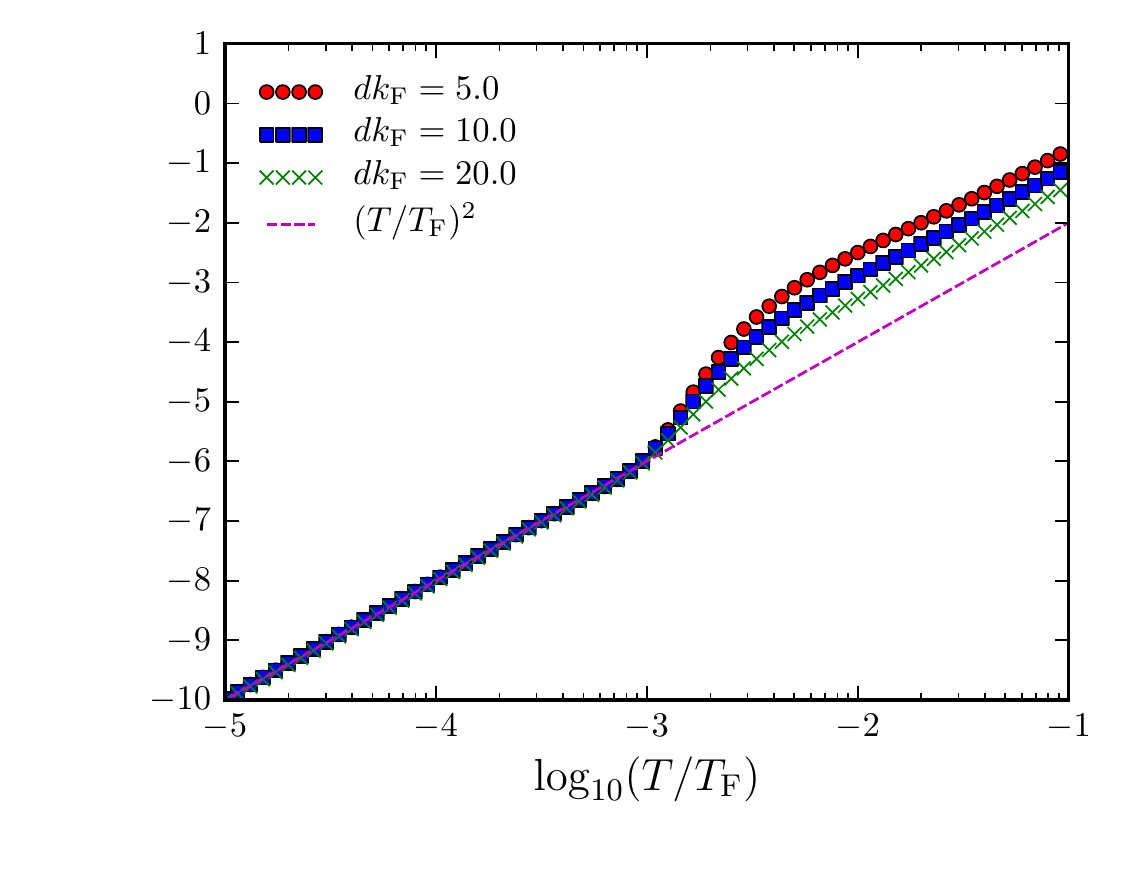}
\caption{(Color online) Top panel: the Coulomb drag transresistivity $\rho^{\rm BT}_{\rm D}$ (in units of $-\rho_0 = - h/e^2$) is plotted as a function of temperature $T$ (in units of the Fermi temperature $T_{\rm F} = \varepsilon_{\rm F}/k_{\rm B}$). Note that both axes are in logarithmic scale. Data in this panel refer to: a carrier density $n = 3.1 \times 10^{14}~{\rm cm}^{-2}$, an interlayer distance $d \approx 2~{\rm nm} $ ($dk_{\rm F} = 5.0$), $\epsilon_1=1$, and $\epsilon_2 = 7.8$. Filled circles label the data obtained by taking into account the frequency dependence $\epsilon_3 = \epsilon_3(\omega)$ reported in Eq.~(\ref{eq:epsilon3omega}). The solid line labels the results obtained by setting $\epsilon_3 = \epsilon_3(\omega=0) =\epsilon_0$ in Eq.~(\ref{eq:dragepsilon3omega}). Bottom panel. 
The Coulomb drag transresistivity $\rho^{\rm BT}_{\rm D}$ (in units of $- \rho_0 a$, where the coefficient $a$ changes with changing $d k_{\rm F}$ -- see text) is plotted as a function of $T/T_{\rm F}$. Note that also in this panel both axes are in logarithmic scale.
Curves labeled by different symbols correspond to different values of the interlayer distance $d$ (density is fixed at the same value used in the top panel, {\it i.e.} $n = 3.1 \times 10^{14}~{\rm cm}^{-2}$). \label{fig:four}}
\end{figure}

The double integral in Eq.~(\ref{eq:dragepsilon3omega}) can be easily performed numerically and some illustrative results are summarized in Fig.~\ref{fig:four}. 
For temperatures $T \lesssim \hbar\omega_0/k_{\rm B}$ we find the usual Fermi-liquid quadratic-in-temperature behavior, $- \rho^{\rm BT}_{\rm D}/\rho_0 = a T^2$ where $\rho_0 = h/e^2$ and 
$a$ is a numerical coefficient whose actual value depends on the carrier density $n$ and on the interlayer distance $d$. However, we clearly see from Fig.~\ref{fig:four} that, for temperatures $T \gtrsim \hbar\omega_0/k_{\rm B}$, the low-temperature Coulomb drag transresistivity deviates from the canonical Fermi-liquid behavior.  

Further deviations from the Fermi-liquid temperature dependence are induced by the explicit (and strong) temperature dependence of the dielectric constant of ${\rm SrTiO}_3$ -- see, for example, Ref.~\onlinecite{couto_prl_2011} -- which here has been neglected for simplicity.

\section{Summary of our main results and conclusions}
\label{sect:conclusions}

In summary, we have presented in a complete and as much pedagogical as possible manner the minimal theory of Coulomb drag 
between two spatially-separated two-dimensional (2D) systems of massless Dirac fermions, which are both away from the charge-neutrality point. 
Our theory relies on second-order perturbation theory in the screened interlayer interaction and on Boltzmann transport theory. 

In this well-established theoretical framework, which has also been adopted by other authors in earlier works~\cite{tse_prb_2007,narozhny_prb_2007,katsnelson_prb_2011,peres_epl_2011,hwang_prb_2011,narozhny_arXiv_2011}, we have clearly demonstrated that the precise functional dependence of the intralayer scattering time on momentum plays absolutely no role in determining  
the low-temperature Coulomb drag transresistivity in the Fermi-liquid regime. This is in contradiction with the findings reported in Refs.~\onlinecite{peres_epl_2011} and~\onlinecite{hwang_prb_2011} while it agrees with the conclusions reached by Narozhny {\it et al.}~\cite{narozhny_arXiv_2011}.

For two layers with identical degeneracies ($N_{{\rm f}, 1} =  N_{{\rm f}, 2} \equiv N_{\rm f}$) and densities ($n_1 = n_2 \equiv n$), we find the following results for the low-temperature Coulomb drag transresistivity: 

1) In the weak-coupling limit -- Eq.~(\ref{eq:rhoD_GG_weakcoupling_final}) -- we find
\be
\lim_{T \to 0}\rho^{\rm BT}_{\rm D} = -\frac{h}{e^2}~\frac{\pi  \zeta (3)}{8}~\frac{\epsilon_2 ^2}{d^4 \alpha^2_{\rm ee}  N^4_{\rm f}k^6_{\rm F}}\frac{k^2_{\rm B}T^2}{\hbar^2 v^2}~;
\ee
The dependence of this result on layer-separation $d$ and doping $n$ agrees with that found in Refs.~\onlinecite{tse_prb_2007,katsnelson_prb_2011} and~\onlinecite{narozhny_arXiv_2011}.

2) In the strong-coupling limit -- Eq.~(\ref{eq:rhoD_GG_strongcoupling_equallayers}) -- we find
\be
\lim_{T \to 0}\rho^{\rm BT}_{\rm D} = - \frac{h}{e^2}~\frac{4 \pi }{3}~\frac{\alpha^2_{\rm ee}}{k^2_{\rm F}}~\frac{k^2_{\rm B} T^2}{\hbar^2 v^2}~{\cal F}(\epsilon_1+\epsilon_3,N_{\rm f}\alpha_{\rm ee})~,
\ee
where the explicit expression of the function ${\cal F}(\epsilon_1+\epsilon_3,N_{\rm f}\alpha_{\rm ee})$ is reported in Eq.~(\ref{eq:Ffunction}). The independence of this result on layer-separation $d$ and the dependence on doping $n$ ($\propto 1/n$) agree with the findings of Ref.~\onlinecite{narozhny_arXiv_2011}, even though the authors of this work captured only the weak-screening $N_{\rm f}\alpha_{\rm ee} \to 0$ asymptotic behavior of the function ${\cal F}(\epsilon_1+\epsilon_3,N_{\rm f}\alpha_{\rm ee})$ and did not take into account the spatial dependence of the dielectric constant along the ${\hat {\bm z}}$ direction.

General results for $n_1 \neq n_2$ and $N_{{\rm f}, 1} \neq N_{{\rm f}, 2}$ can be found in Eqs.~(\ref{eq:rhoD_GG_weakcoupling_final}) and~(\ref{eq:rhoD_GG_strongcoupling_equallayers}).

Finally, we have shown how deviations from the Fermi-liquid temperature dependence can occur when one of dielectric constants of the media surrounding the Dirac-fermion layers depends strongly on frequency and temperature. Double-layer graphene systems deposited, for example, on ${\rm SrTiO}_3$, a very well-known insulator close to a ferroelectric instability~\cite{couto_prl_2011,han_apl_2007}, can be used as a testbed for this idea.

In the future, we plan to present a systematic numerical study of Eq.~(\ref{eq:rho_general_finiteT}) at arbitrary temperatures and to investigate more deeply the strong-coupling limit by studying i) the impact of diagrams of order higher than two in the screened interlayer interaction and ii) beyond-RPA corrections.

\acknowledgments
We wish to thank Andre Geim, Allan MacDonald, and Misha Titov for very useful and stimulating discussions. Work in Pisa was supported by the Italian Ministry of Education, University, and Research (MIUR) through the program ``FIRB - Futuro in Ricerca 2010" Grant No. RBFR10M5BT (``PLASMOGRAPH: plasmons and terahertz devices in graphene"). T.T. and M.I.K. acknowledge financial support from the 
Stichting voor Fundamenteel Onderzoek der Materie (FOM) (The Netherlands).

\appendix

\section{Boltzmann transport approach to the drag conductivity}
\label{appendix:boltzmann}

The Boltzmann transport equation for the distribution function $f_\ell({\bm k}, \lambda)$ of the $\ell$-th layer reads:
\be\label{eq:boltzmann}
\partial_t f_\ell({\bm k}, \lambda) + \nabla_{\bm k}f_\ell ({\bm k}, \lambda) \cdot \partial_t {\bm k} = I_\ell({\bm k}, \lambda) + \gamma_{\ell, {\bar \ell}}({\bm k},\lambda)~,
\ee
where $I_\ell({\bm k}, \lambda)$ is the so-called collision integral, which describes intralayer scattering events, while the term $\gamma_{\ell, {\bar \ell}}({\bm k},\lambda)$ describes momentum transfer between the two Dirac-fermion layers (${\bar \ell} = 2$ if $\ell =1$ and ${\bar \ell}=1$ if $\ell = 2$) and is obviously absolutely crucial in the Coulomb drag problem. We are interested in the steady-state regime in which 
$\partial_t  f_\ell({\bm k}, \lambda) =0$ and employ a ``generalized relaxation time approximation" for the intralayer collision integral, {\it i.e.}
\be
I_\ell({\bm k}, \lambda) = - \frac{f_\ell({\bm k}, \lambda) - f^{(0)}_\ell({\bm k}, \lambda)}{\tau_\ell({\bm k})}~,
\ee
where $f^{(0)}_\ell({\bm k}, \lambda) = n_{\rm F}(\xi^{(\ell)}_{{\bm k}, \lambda})$ is the equilibrium distribution function. Below we will assume that the transport scattering time $\tau_\ell({\bm k})$ is isotropic.

The interlayer collision integral reads
\begin{widetext}
\ber
\gamma_{\ell, {\bar \ell}}({\bm k}, \lambda) &=& 2\pi N_{{\rm f}, {\bar \ell}} \sum_{\lambda', \lambda'', \lambda '''} \int \frac{d^2 {\bm q}}{(2\pi)^2} \int \frac{d^2 {\bm k}'}{(2\pi)^2}  |U_{12} (q, \xi^{(\ell)}_{{\bm k} , \lambda } - \xi^{(\ell)} _{{\bm k} + {\bm q} , \lambda'} )|^2 \delta (\xi^{(\ell)}_{{\bm k} , \lambda} - \xi^{(\ell)}_{{\bm k} + {\bm q}, \lambda'} + \xi^{({\bar\ell})}_{{\bm k}' , \lambda'' } 
- \xi^{({\bar \ell})}_{{\bm k} ' - {\bm q} , \lambda'''} ) 
\nonumber\\
&\times&
\Big\{f_{\ell}({\bm k} , \lambda) [1-f_\ell({\bm k} + {\bm q} , \lambda')] f_{\bar \ell} ({\bm k}' , \lambda '') [1-f_{\bar \ell} ({\bm k}' - {\bm q}, \lambda''')] \nonumber\\
&-& [1-f_\ell({\bm k} , \lambda)] f_\ell ({\bm k} + {\bm q}, \lambda')[1 -f_{\bar \ell}({\bm k}' , \lambda '')] f_{\bar \ell}({\bm k}' - {\bm q}, \lambda''')\Big\}\nonumber\\
&\times&\frac{1 + \lambda \lambda ' \cos (\varphi _{{\bm k}} - \varphi_{{\bm k } + {\bm q}})}{2} \frac{1 + \lambda'' \lambda ''' \cos ( \varphi_{{\bm k}'} - \varphi_{{\bm k}' - {\bm q}})}{2}~.
\eer
\end{widetext}
The degeneracy factor $N_{{\rm f}, {\bar \ell}}$ can be understood from a careful analysis of the conservation of spin and valley degrees of freedom during the scattering process.

In the semiclassical limit $\partial_t {\bm k} = - e {\bm E}_\ell$, where ${\bm E}_\ell$ is the electric field in the $\ell$-th layer. 
In the regime of small electric fields we can linearize the Boltzmann transport equation by writing
\be
f_\ell({\bm k},\lambda) = f^{(0)}_\ell({\bm k},\lambda) + f^{(1)}_\ell({\bm k},\lambda)~.
\ee
The linearized Boltzmann equation in the steady-state regime reads
\be\label{eq:boltzmannlinerized}
- e \nabla_{\bm k}f^{(0)}_\ell ({\bm k}, \lambda) \cdot {\bm E}_\ell = -\frac{f^{(1)}_\ell({\bm k},\lambda)}{\tau_\ell({\bm k})} + \gamma^{({\rm linear})}_{\ell, {\bar \ell}}({\bm k}, \lambda)~,
\ee
where the linearized interlayer collision integral reads
\begin{widetext}
\ber\label{eq:linearizedinterlayercollisions}
\gamma^{({\rm linear})}_{\ell, {\bar \ell}}({\bm k}, \lambda) &=& 2\pi N_{{\rm f}, {\bar \ell}} \sum_{\lambda', \lambda'', \lambda '''} \int \frac{d^2 {\bm q}}{(2\pi)^2} \int \frac{d^2 {\bm k}'}{(2\pi)^2}  |U_{12} (q, \xi^{(\ell)}_{{\bm k} , \lambda } - \xi^{(\ell)} _{{\bm k} + {\bm q} , \lambda'} )|^2 \delta (\xi^{(\ell)}_{{\bm k} , \lambda} - \xi^{(\ell)}_{{\bm k} + {\bm q}, \lambda'} + \xi^{({\bar\ell})}_{{\bm k}' , \lambda'' } 
- \xi^{({\bar \ell})}_{{\bm k} ' - {\bm q} , \lambda'''} ) 
\nonumber\\
&\times&
\Big\{
f^{(1)}_\ell({\bm k} , \lambda) f^{(0)}_{\bar \ell}({\bm k}' , \lambda '') [1 - f^{(0)}_\ell({\bm k} + {\bm q} , \lambda')] [1-f^{(0)}_{\bar \ell}({\bm k}' - {\bm q}, \lambda''')]
\nonumber\\
&-&
f^{(1)}_\ell({\bm k} + {\bm q} , \lambda') f^{(0)}_\ell({\bm k} , \lambda) f^{(0)}_{\bar \ell}({\bm k}' , \lambda '') [1-f^{(0)}_{\bar \ell}({\bm k}' - {\bm q}, \lambda''')]
\nonumber\\
&+&
f^{(1)}_{\bar \ell}({\bm k}' , \lambda '') f^{(0)}_\ell({\bm k} , \lambda) [1 - f^{(0)}_\ell({\bm k} + {\bm q} , \lambda')] [1-f^{(0)}_{\bar \ell}({\bm k}' - {\bm q}, \lambda''')]
\nonumber\\
&-&
f^{(1)}_{\bar \ell}({\bm k}' - {\bm q}, \lambda''') f^{(0)}_\ell({\bm k} , \lambda) f^{(0)}_{\bar \ell}({\bm k}' , \lambda '') [1 - f^{(0)}_\ell({\bm k} + {\bm q} , \lambda')]
\nonumber\\
&-&
f^{(1)}_\ell({\bm k} , \lambda) f^{(0)}_\ell({\bm k} + {\bm q} , \lambda') f^{(0)}_{\bar \ell}({\bm k}' - {\bm q}, \lambda''') [1 - f^{(0)}_{\bar \ell}({\bm k}' , \lambda '')]
\nonumber\\
&+&
f^{(1)}_\ell({\bm k} + {\bm q} , \lambda') f^{(0)}_{\bar \ell}({\bm k}' - {\bm q}, \lambda''') [1 - f^{(0)}_\ell({\bm k} , \lambda)] [1 - f^{(0)}_{\bar \ell}({\bm k}' , \lambda '')] 
\nonumber\\
&-&
f^{(1)}_{\bar \ell}({\bm k}' , \lambda '') f^{(0)}_\ell({\bm k} + {\bm q} , \lambda') f^{(0)}_{\bar \ell}({\bm k}' - {\bm q}, \lambda''') [1 - f^{(0)}_\ell({\bm k} , \lambda)] 
\nonumber\\
&+&
f^{(1)}_{\bar \ell}({\bm k}' - {\bm q}, \lambda''')f^{(0)}_\ell({\bm k} + {\bm q} , \lambda') [1 - f^{(0)}_\ell({\bm k} , \lambda)]  [1 - f^{(0)}_{\bar \ell}({\bm k}' , \lambda '')]
\Big\}\nonumber\\
&\times&\frac{1 + \lambda \lambda ' \cos (\varphi _{{\bm k}} - \varphi_{{\bm k } + {\bm q}})}{2} \frac{1 + \lambda'' \lambda ''' \cos ( \varphi_{{\bm k}'} - \varphi_{{\bm k}' - {\bm q}})}{2}~.
\eer
\end{widetext}
We now seek solutions of Eqs.~(\ref{eq:boltzmannlinerized})-(\ref{eq:linearizedinterlayercollisions}) of the form
\ber
f^{(1)}_\ell({\bm k},\lambda) &=& \psi^{({\rm intra})}_\ell(k,\lambda) {\hat {\bm k}} \cdot {\bm E}_\ell \nonumber\\
&+& \psi^{({\rm inter})}_\ell(k,\lambda) {\hat {\bm k}} \cdot {\bm E}_{\bar \ell}~.
\eer
In the following we will assume that $|\psi^{({\rm intra})}_\ell(k,\lambda)| \gg |\psi^{({\rm inter})}_\ell(k,\lambda)|$. The coefficients $\psi^{({\rm intra})}_\ell(k,\lambda)$ and $\psi^{({\rm inter})}_\ell(k,\lambda)$ can be found by calculating the functional derivative of the linearized 
Boltzmann transport equation with respect to ${\bm E}_\ell$ and ${\bm E}_{\bar \ell}$, respectively, {\it i.e.}
\be\label{eq:derivativewrtEelle}
- e \nabla_{\bm k}f^{(0)}_\ell ({\bm k}, \lambda) \simeq - {\hat {\bm k}} \frac{\psi^{({\rm intra})}_\ell(k,\lambda)}{\tau_\ell({\bm k})} 
\ee
and
\be\label{eq:derivativewrtEellebar}
0 = -{\hat {\bm k}} \frac{\psi^{({\rm inter})}_\ell(k,\lambda)}{\tau_\ell({\bm k})} + \frac{\delta \gamma^{({\rm linear})}_{\ell, {\bar \ell}}({\bm k}, \lambda)}{\delta {\bm E}_{\bar \ell}}~.
\ee

In writing Eq.~(\ref{eq:derivativewrtEelle}) we have neglected the small term $\delta \gamma^{({\rm linear})}_{\ell, {\bar \ell}}({\bm k}, \lambda)/\delta {\bm E}_{\ell}$. The solution of Eq.~(\ref{eq:derivativewrtEelle}) reads
\be \label{psi_intra}
\psi^{({\rm intra})}_\ell(k,\lambda) = e {\hat {\bm k}} \cdot \nabla_{\bm k} \xi^{(\ell)}_{{\bm k}, \lambda} \frac{\partial f^{(0)}_\ell({\bm k}, \lambda)}{\partial \xi^{(\ell)}_{{\bm k}, \lambda}}\tau_\ell({\bm k})~.
\ee

The second term on the r.h.s. of Eq.~(\ref{eq:derivativewrtEellebar}) can be calculated from Eq.~(\ref{eq:linearizedinterlayercollisions}). Solving Eq.~(\ref{eq:derivativewrtEellebar}) we find
\begin{widetext}
\ber\label{eq:psi_inter}
\psi^{({\rm inter})}_\ell(k,\lambda) &=& -2\pi e \beta \tau_\ell({\bm k}) N_{{\rm f}, {\bar \ell}} \sum_{\lambda', \lambda'', \lambda '''} \int \frac{d^2 {\bm q}}{(2\pi)^2} \int \frac{d^2 {\bm k}'}{(2\pi)^2}  |U_{12} (q, \xi^{(\ell)}_{{\bm k} , \lambda } - \xi^{(\ell)} _{{\bm k} + {\bm q} , \lambda'} )|^2
\nonumber\\
&\times&
\delta (\xi^{(\ell)}_{{\bm k} , \lambda} - \xi^{(\ell)}_{{\bm k} + {\bm q}, \lambda'} + \xi^{({\bar\ell})}_{{\bm k}' , \lambda'' } 
- \xi^{({\bar \ell})}_{{\bm k} ' - {\bm q} , \lambda'''} ) 
\Big\{f^{(0)}_\ell({\bm k} , \lambda)f^{(0)}_{\bar \ell}({\bm k}' , \lambda '') [1 - f^{(0)}_\ell({\bm k} + {\bm q} , \lambda')] [1-f^{(0)}_{\bar \ell}({\bm k}' - {\bm q}, \lambda''')]\Big\}
\nonumber\\
&\times&
\Big[\tau_{\bar \ell} ({\bm k}') \nabla_{{\bm k}'}\xi^{({\bar \ell})}_{{\bm k}' , \lambda''}  - \tau_{\bar \ell}({\bm k}' - {\bm q})\nabla_{{\bm k}'}\xi^{({\bar \ell})}_{{\bm k}' -{\bm q} , \lambda'''}\Big]\cdot {\hat {\bm k}} \nonumber\\
&\times&\frac{1 + \lambda \lambda ' \cos (\varphi _{{\bm k}} - \varphi_{{\bm k } + {\bm q}})}{2} \frac{1 + \lambda'' \lambda ''' \cos ( \varphi_{{\bm k}'} - \varphi_{{\bm k}' - {\bm q}})}{2}~.
\eer
\end{widetext}

To obtain Eq.~(\ref{eq:psi_inter}) we used the identities 
\begin{equation}
\frac{\partial f^{(0)}_\ell({\bm k}, \lambda)}{\partial \xi^{(\ell)}_{{\bm k}, \lambda}} = \beta f^{(0)}_\ell({\bm k}, \lambda)
[1 - f^{(0)}_\ell({\bm k}, \lambda)]
\end{equation}
and
\begin{equation}
f^{(0)}_\ell({\bm k}, \lambda) [1 - f^{(0)}_\ell({\bm k}', \lambda')] = \frac{f^{(0)}_\ell({\bm k}, \lambda) - f^{(0)}_\ell({\bm k}', \lambda')}{1 - \exp[\beta(\xi^{(\ell)}_{{\bm k}, \lambda} - \xi^{(\ell)}_{{\bm k}', \lambda'})]}
~.
\end{equation}

The current density in the $\ell$-th layer reads
\ber
{\bm J}_\ell &=& -e N_{{\rm f}, \ell} \sum_\lambda \int \frac{d^2{\bm k}}{(2\pi)^2} f^{(1)}_\ell({\bm k},\lambda) \nabla_{\bm k} \xi^{(\ell)}_{{\bm k}, \lambda}
\nonumber\\
&\equiv& {\bm J}^{({\rm intra})}_\ell + {\bm J}^{({\rm inter})}_\ell ~.
\eer
Using Eq.~(\ref{psi_intra}) we find ${\bm J}^{({\rm intra})}_\ell$:
\ber \label{current_density_intra}
{\bm J}^{({\rm intra})}_\ell &=& e^2 N_{{\rm f}, \ell} \sum_\lambda \int \frac{d^2{\bm k}}{(2\pi)^2}
\tau_\ell({\bm k}) \big[{\hat {\bm k}} \cdot \nabla_{\bm k} \xi^{(\ell)}_{{\bm k}, \lambda}\big] ({\hat {\bm k}}\cdot {\bm E}_\ell) \nonumber\\
&\times& \Bigg[-\frac{\partial f^{(0)}_\ell({\bm k}, \lambda)}{\partial \xi^{(\ell)}_{{\bm k}, \lambda}}\Bigg]
\nabla_{\bm k} \xi^{(\ell)}_{{\bm k}, \lambda}
\nonumber\\
&=& {\bm E}_\ell \frac{e^2 N_{{\rm f}, \ell} }{2} \sum_\lambda \int \frac{d^2{\bm k}}{(2\pi)^2}
\tau_\ell({\bm k}) \Bigg( \frac{\partial \xi^{(\ell)}_{{\bm k}, \lambda}}{\partial k} \Bigg)^2
\nonumber\\
&\times& \Bigg[-\frac{\partial f^{(0)}_\ell({\bm k}, \lambda)}{\partial \xi^{(\ell)}_{{\bm k}, \lambda}}\Bigg]
~.
\eer
In the last equality of Eq.~(\ref{current_density_intra}) we used the fact that
\be \label{eq:nabla_k_xi_app_A}
\nabla_{\bm k} \xi^{(\ell)}_{{\bm k}, \lambda} = {\hat {\bm k}} \frac{\partial \xi^{(\ell)}_{{\bm k}, \lambda}}{\partial k}
~,
\ee
(since $\xi^{(\ell)}_{{\bm k}, \lambda}$ depends only on $|{\bm k}|$) and that the average over the angle of $({\hat {\bm k}}\cdot {\bm v}) {\hat {\bm k}} = {\bm v}/2$ for any constant vector ${\bm v}$ (in $D=2$ spatial dimensions). From Eq.~(\ref{current_density_intra}) we find immediately the longitudinal intralayer conductivity
\ber \label{eq:sigma_ell_app_A}
\sigma_\ell &\equiv& \frac{\delta J^{(i)}_\ell}{\delta E^{(i)}_\ell} = 
\frac{e^2 N_{{\rm f}, \ell} }{2} \sum_\lambda \int \frac{d^2{\bm k}}{(2\pi)^2}
\tau_\ell({\bm k}) \Bigg( \frac{\partial \xi^{(\ell)}_{{\bm k}, \lambda}}{\partial k} \Bigg)^2
\nonumber\\
&\times& \Bigg[-\frac{\partial f^{(0)}_\ell({\bm k}, \lambda)}{\partial \xi^{(\ell)}_{{\bm k}, \lambda}}\Bigg]
\nonumber\\
&=&
\frac{e^2 v^2 N_{{\rm f}, \ell} }{2} \sum_\lambda \int \frac{d^2{\bm k}}{(2\pi)^2}
\tau_\ell({\bm k}) \Bigg[-\frac{\partial f^{(0)}_\ell({\bm k}, \lambda)}{\partial \xi^{(\ell)}_{{\bm k}, \lambda}}\Bigg]
~,
\nonumber\\
\eer
where $i$ is a Cartesian index and in the last equality we used the fact that $\xi^{(\ell)}_{{\bm k}, \ell} = \lambda v |{\bm k}| -\mu_\ell$. Note that Eq.~(\ref{eq:sigma_ell_app_A}) coincides with Eq.~(\ref{eq:cond_intra_finiteT}) in the main text.

The interlayer current density reads
\be \label{current_density_inter}
{\bm J}^{({\rm inter})}_\ell = -e N_{{\rm f}, \ell}  \sum_\lambda \int \frac{d^2{\bm k}}{(2\pi)^2}
\psi^{({\rm inter})}(k,\lambda) ({\bm k}\cdot {\bm E}_{\bar \ell})
\nabla_{\bm k} \xi^{(\ell)}_{{\bm k}, \lambda}
~,
\ee
where $\psi^{({\rm inter})}(k,\lambda)$ is given in Eq.~(\ref{eq:psi_inter}). After some straightforward algebra we obtain
\begin{widetext}
\ber\label{eq:current_density_inter_final}
{\bm J}^{({\rm inter})}_\ell &=& {\bm E}_{\bar \ell} \frac{2\pi N_{{\rm f}, \ell} N_{{\rm f}, {\bar \ell}} \beta e^2}{2}  \sum_{\lambda , \lambda ' , \lambda '' , \lambda'''}\int \frac{d^2{\bm q}}{(2\pi)^2} \int \frac{d^2 {\bm k}}{(2\pi)^2} \int \frac{d^2 {\bm k}'}{(2\pi)^2} 
|U_{1 2 } ({\bm q}, \xi^{(\ell)}_{{\bm k} , \lambda } - \xi^{(\ell)} _{{\bm k} + {\bm q} , \lambda'} )|^2
\nonumber\\
&\times&
\delta (\xi^{(\ell)}_{{\bm k} , \lambda} - \xi^{(\ell)}_{{\bm k} + {\bm q}, \lambda'} + \xi^{({\bar \ell})}_{{\bm k}' , \lambda'' } - \xi^{({\bar \ell})}_{{\bm k} ' -{\bm q} , \lambda '''} )
\Big\{ f_\ell^{(0)} ({\bm k} , \lambda)f_{\bar \ell}^{(0)} ({\bm k}' , \lambda '')  [1 - f_\ell^{(0)} ({\bm k} + {\bm q} , \lambda')] [1-f_{\bar\ell}^{(0)} ({\bm k}' - {\bm q}, \lambda''')] \Big\}
\nonumber\\
&\times&
\left[\tau_\ell ({\bm k} ) \nabla_{\bm k} \xi^{(\ell)}_{{\bm k} , \lambda}  - \tau_\ell ({\bm k} + {\bm q}) \nabla_{\bm k} \xi^{(\ell)}_{{\bm k} +{\bm q} , \lambda'} \right]
\cdot
\left[\tau_{\bar \ell} ({\bm k}' ) \nabla_{{\bm k}'} \xi^{({\bar \ell})}_{{\bm k}' , \lambda''} - \tau_{\bar \ell} ({\bm k}' - {\bm q}) \nabla_{{\bm k}'}\xi^{({\bar \ell})}_{{\bm k}' -{\bm q} , \lambda'''} \right]
\nonumber\\
&\times&
\frac{1 + \lambda \lambda ' \cos (\varphi _{{\bm k}} - \varphi_{{\bm k } + {\bm q}})}{2} \frac{1 + \lambda'' \lambda ''' \cos ( \varphi_{{\bm k}'} - \varphi_{{\bm k}' - {\bm q}})}{2}~.
\nonumber\\
\eer
\end{widetext}
Notice that Eq.~(\ref{eq:current_density_inter_final}) has been recasted in a symmetric form by means of a simultaneous exchange of dummy variables (${\bm k} \leftrightarrow {\bm k}+{\bm q}$, ${\bm k}' \leftrightarrow {\bm k}'-{\bm q}$, $\lambda \leftrightarrow \lambda'$, and $\lambda'' \leftrightarrow \lambda'''$) and thus the integral is multiplied by an additional factor $1/2$. From Eq.~(\ref{eq:current_density_inter_final}) we can derive the longitudinal drag conductivity $\sigma_{\rm D}$, which reads as following:
\begin{widetext}
\be\label{eq:sigma_D_boltzmann_app_A}
\sigma_{\rm D} \equiv \frac{\delta J^{(i)}_\ell}{\delta E^{(i)}_{\bar \ell}} = \frac{\beta e^2}{16\pi} \int \frac{d^2 {\bm q}}{(2\pi)^2} \int_{-\infty}^{+\infty} d \omega \frac{|U_{12} (q,\omega)|^2}{\sinh^2(\beta \omega/2)} \Gamma^{\rm BT}_1({\bm q},\omega) \Gamma^{\rm BT}_2({\bm q}, \omega) ~,
\ee
\end{widetext}
where the function $\Gamma^{\rm BT}_\ell({\bm q},\omega)$ coincides with the one defined in Eq.~(\ref{eq:gamma_pol2}) of the main text. 
In writing Eq.~(\ref{eq:sigma_D_boltzmann_app_A}) we have introduced an auxiliary variable $\omega$ to disentangle ${\bm k}$ from ${\bm k}'$:
\be
\delta (\xi - \xi') = \int d \omega \delta (\xi - \omega) \delta (\xi' - \omega) ~.
\ee
\end{document}